\documentclass[journal]{IEEEtran}

\AtBeginDocument{%
  \providecommand\BibTeX{{%
    \normalfont B\kern-0.5em{\scshape i\kern-0.25em b}\kern-0.8em\TeX}}}

\usepackage{graphicx}
\usepackage{threeparttablex}
\usepackage{multirow}
\usepackage[lofdepth,lotdepth]{subfig}
\usepackage{hyperref}
\usepackage{pifont}
\usepackage{amssymb}

\newcommand*\rot{\rotatebox{90}}

\title{Vulnerability and Transaction behavior based detection of Malicious Smart Contracts}

\author{
\IEEEauthorblockN{Rachit Agarwal, Tanmay Thapliyal, Sandeep Kumar Shukla}\\
\IEEEauthorblockA{
 CSE Department, IIT Kanpur\\
 Email: \{rachitag, tanmayt, sandeeps\}@iitk.ac.in}
}
\begin{document}

\maketitle

\begin{abstract} 

Smart Contracts (SCs) in Ethereum can automate tasks and provide different functionalities to a user. Such automation is enabled by the `Turing-complete' nature of the programming language (Solidity) in which SCs are written. This also opens up different vulnerabilities and bugs in SCs that malicious actors exploit to carry out malicious or illegal activities on the cryptocurrency platform. In this work, we study the correlation between malicious activities and the vulnerabilities present in SCs and find that some malicious activities are correlated with certain types of vulnerabilities. We then develop and study the feasibility of a scoring mechanism that corresponds to the severity of the vulnerabilities present in SCs to determine if it is a relevant feature to identify suspicious SCs. We analyze the utility of severity score towards detection of suspicious SCs using unsupervised machine learning (ML) algorithms across different temporal granularities and identify behavioral changes. In our experiments with on-chain SCs, we were able to find a total of 1094  benign SCs across different granularities which behave similar to malicious SCs, with the inclusion of the smart contract vulnerability scores in the feature set.
\end{abstract}


\section{Introduction}\label{sec:intro}

Ethereum was the first blockchain platform to enable programming Turing-complete smart contracts (SCs). However, the use of SCs in such expressible language also opens the doors to vulnerabilities and bugs. These enabled various participants in the Ethereum platform (organizations and individuals) to exploit the vulnerabilities for malicious activities (such as Bitpoint Hacks). An Ethereum account is \textit{malicious} if it performs, facilitates, or is suspected to be involved in different illegal activities such as \textit{Phishing}, \textit{Gambling}, and \textit{Ponzi} schemes. While these malicious activities are often socially motivated (for example, Gambling and Phishing) and do not exploit and SC vulnerabilities, in several other cases (for example, Lendf Hack and Akropolis hack\footnote{https://www.zdnet.com/article/hacker-steals-2-million-from-cryptocurrency-service-akropolis/}), malicious activities are carried out due to  the exploitation of the bugs and vulnerabilities .

In~\cite{Alkhalifah2020}, the authors survey different vulnerabilities that exist in an SC. While bugs and vulnerabilities are the main focus of their survey, it lacks behavioral understanding as it does not consider the transactions performed by an SC. In Ethereum, there are two types of transactions performed by an SC: internal and external. External transactions are recorded on the ledger, while the internal transactions are not recorded on the ledger but can be obtained using Ethereum Virtual Machines (EVM). Internal transactions are mainly of 5 types: CALL, CALLCODE, SUICIDE, DELEGATECALL, and  CREATE. Briefly, a CALL transaction refers to a transaction where an SC invokes another SC. In a CALLCODE and DELEGATECALL, the caller calls an SC on behalf of another SC. Note, the DELEGATECALL opcode is a newer version of the CALLCODE opcode. SUICIDE opcode allows an SC to self-destruct, causing all the SC's internal transactions to be lost. At the same time, a CREATE opcode enables an SC to create a new SC. Due to such types of opcodes, a malicious SC could create multiple SCs to evade detection. As identified in~\cite{Durieux2020}, irrespective of whether they are malicious or not, most of the created  SCs have the same code and therefore have the same vulnerabilities. Nonetheless, only a fraction of SCs that have vulnerabilities are exploited~\cite{Perez2021}. SCs that have similar code to those exploited are not detected/marked as malicious as those SCs themselves are not exploited.

Thus, we ask if \textit{we can train a machine learning (ML) algorithm to detect an SC that shows malicious behavior (even when the SC is not marked as malicious) and has vulnerabilities}. Note that when an SC is not marked malicious, we can only consider it as a \textit{suspect} and cannot officially label it as malicious. Various state-of-the-art approaches exist that detect malicious and vulnerable SCs, such as~\cite{Durieux2020, agarwal2020detecting}. In~\cite{agarwal2020detecting}, the authors consider SCs as regular accounts (or the Externally Owned Accounts (EOA)s) and neglect SC vulnerabilities and their internal transactions. Further, as malicious activities can be of different types, the approach does not correlate vulnerabilities exploited by a particular malicious activity. In~\cite{Durieux2020}, the authors use various SC code analysis tools in the context of the Decentralized Application Security Project (DASP)\footnote{\url{https://dasp.co/}} and identified the top 10 vulnerabilities. Nonetheless, more than 36 (still growing) vulnerabilities are identified under Smart contract Weakness Classification (SWC)\footnote{\url{https://swcregistry.io/}}, some of which are not considered in prior work. 

Thus, we are motivated to answer: \textbf{\textit{(Q1)}} is there a correlation between a particular malicious activity and a vulnerability in the SC, and if so, does the severity of a vulnerability correspond to its exploitability in committing malicious activities? \textbf{\textit{(Q2)}} is the vulnerability severity score an important feature to learn by an ML algorithm that aims at detecting malicious accounts, and should it be used to identify malicious SCs (or the suspects)? And \textbf{\textit{(Q3)}} do the SCs that are not marked malicious also show malicious behavior in different temporal granularities (observations in different temporal scales), and can the usage of severity score as a feature along with other temporal features detect such SCs? 

To answer these questions, we first use different SC vulnerability analysis tools to analyze SCs. In parallel, we generate their transaction-based, graph-based, and temporal-based features from both internal and external transaction data available via Etherscan~\cite{etherscanApi}. The list of malicious SCs is available using~\cite{etherscanLC}. As this list is limited, we develop a CREATE transaction-based graph to identify all those SCs that a malicious SC creates, assuming that the parent and child of the malicious SC will be malicious. Of course, we understand that, in reality, this assumption might fail due to transaction behavior. Our analysis reveals that for a subset of SC vulnerabilities, such as those under \textit{CWE-841}, there exists a correlation between them and the transaction behavior shown by SCs. Motivated by this, we include severity score as a feature and apply different ML algorithms to answer whether the inclusion of such feature improves the detection of malicious SCs. Towards this, we create two different datasets, one that includes both severity and transaction-based features and the other, which has only transaction-based features. We analyze them and check if better silhouette scores are obtained when compared with the silhouette scores when the severity score was not included in the feature set. Our analysis reveals that not all SCs with similar vulnerabilities cluster together. In the process, we also validate the findings of~\cite{Perez2021} and observe that the existence of a `vulnerability' does not imply that an SC is exploited. Further, we divide the dataset into different temporal granularities such as Daywise (1-Day), 3-Day, and month (1-Month). On the identified sub-datasets, we recompute the features and apply the unsupervised ML algorithm to identify the probability of a benign SC being malicious. We observe that SCs do show malicious behavior in different temporal granularities. For instance, when observing the behavior of SCs in the 3-Day granularity, we find 24 benign SCs that behave similar to the malicious SCs over time, and thus, we consider them to be suspects. We also observe a difference in the number of benign SCs that behave similar to malicious SCs in a particular granularity when we include the severity score as a feature. For instance, when we use both transaction and severity-based features in the Month-wise granularity, we discover 1066 suspects compared to 866 suspects when we consider only transaction-based features.

In summary our core contributions are:
\begin{itemize}
    \item We present a \textit{\textbf{mapping between different vulnerability vocabularies in the domain of SC vulnerabilities}}. This mapping is based on existing vulnerabilities present in the deployed SCs. It provides a clear understanding of multiple names with which a particular vulnerability is referred to by the SC code analysis tools. 
    
    \item We validate that \textbf{\textit{all the SCs with vulnerabilities are not usually exploited}}, and the \textit{\textbf{severity scores of the vulnerabilities do not impact the transaction behavior}}. Our findings are based on the currently known ground truth about the SCs. Nonetheless, \textit{\textbf{there exists a correlation between the type of malicious activity and the vulnerability}}. For example, CWE-362 vulnerability is only present in the SCs related to Phishing schemes amongst the malicious class.
    
    \item Using our methodology, we \textit{\textbf{identify 2 SCs as potential suspects}} using the \textit{\textbf{K-Means}} algorithm. K-Means performed the best among the set of different unsupervised ML algorithms. Note that the behavior of SCs could change across different temporal granularities. An analysis across the temporal granularities reveals 892 SCs (866 in 1-Month + 24 in 3-Day + 2 in 1-Day) as potential suspects when only transaction-based features are used. This number changes to 1094 SCs (1066 in 1-Month + 24 in 3-Day + 4 in 1-Day) as potential suspects when both transaction-based features and severity score-based features are used.
\end{itemize}

In the rest of the paper, in section~\ref{sec:rw}, we present an overview of the state-of-the-art techniques used to detect accounts behaving maliciously in blockchains. In sections~\ref{sec:method}, we present a detailed description of our methodology. This is followed by an in-depth evaluation along with the results in section~\ref{sec:eval}. We finally conclude in section~\ref{sec:conclusion} providing details on prospective future work.
 
\section{Background and Related Work}\label{sec:rw}

Various studies focus on the detection of malicious activities in the blockchain. While some focus on detecting vulnerabilities in the SCs, others analyze the blockchain by observing the transaction-based features or using both transactions and source code-based features. Note that very few focus on studying the impact of existing vulnerabilities and transaction behavior in determining suspects. In the following subsections, we briefly survey the vulnerabilities detected by the tools and approaches that use transactions to classify or cluster malicious accounts.

\subsection{Vulnerability Detection}\label{sec:CodeAnalT}

Different vocabularies exist which classify various SC vulnerabilities.  In~\cite{Dingman2019}, the authors use the NIST bug framework\footnote{\url{https://samate.nist.gov/BF/}} to categorize different SC vulnerabilities into four categories: \textit{Security}, \textit{Operational}, \textit{Functional}, and \textit{Developmental}. Similarly, in~\cite{Durieux2020}, the authors map different SC vulnerabilities to the top 10 DASP identified vulnerabilities. We provide a short description of these ten different types of SC vulnerabilities in Appendix~\ref{app:d10}. Nonetheless, specific to vulnerabilities present in the SCs, the Smart contract Weakness Classification (SWC) vocabulary exists. Although generated from Common Weakness Enumeration (CWE: a broader classification nomenclature for vulnerabilities), SWC is yet to be standardized and, in some cases, does not cover all the vulnerabilities. For the sake of completeness, in Table~\ref{table:vulnerabilityMapping}, we present the relation between these vocabularies barring the NIST bug framework as there are pending updates to the framework. We find that as these vocabularies are not standard, the interpretation of a vulnerability and its severity varies. For severity scores, we obtain them from individual code analysis tools, and in case of a clash, we chose an interpretation that has a higher severity. Note that Table~\ref{table:vulnerabilityMapping} lists only those vulnerabilities that are present in all the SCs in our dataset. Here we also note that \textit{Bad Randomness} and \textit{Front Running} vulnerabilities defined in DASP are not present in any SC. Further, if a vulnerability is not present in some vocabulary, we mark the corresponding cell in the table with a ``$-$''. We put in our best effort to minimize the number of ``$-$'' and underline those we infer.

\begin{table*}
\center
\caption{Vulnerabilities}\label{table:vulnerabilityMapping}
\begin{tabular}{|l|l|l|l|l||l|l|l|l|l|}
\rot{\textbf{Severity}} & {{\textbf{Vulnerability}}} & \rot{\textbf{DASP-10~\cite{Durieux2020}}} & \rot{\textbf{SWC}} & \rot{\textbf{CWE}}&\rot{\textbf{Severity}} & {{\textbf{Vulnerability}}} & \rot{\textbf{DASP-10~\cite{Durieux2020}}} & \rot{\textbf{SWC}} & \rot{\textbf{CWE}}\\
\hline
 
H & Arbitrary-send$\dagger$ & \multirow{5}{*}{\rot{Acc. Control}} & 124 & 123 & H & Uninitialized state$\dagger$ & \multirow{26}{*}{\rot{Unknown}} & 109 & 824 \\ \cline{1-2} \cline{4-7}  \cline{9-10}
H & Ether send $\ddagger$ &  & 105 & 284 & H & Uninitialized storage$\dagger$ &  & 109 & 824 \\ \cline{1-2} \cline{4-7}  \cline{9-10}
H & Unprotected self destruct$\dagger\ddagger\Diamond$ &  & 106 & 284 & H & Shadowing state$\dagger$ &  & 119 & 710 \\ \cline{1-2} \cline{4-7}  \cline{9-10}
H & Delegate call$\dagger\ddagger$ &  & 112 & 829 & H & Locked Ether$\triangleleft\dagger$ &  & - & -\\ \cline{1-2} \cline{4-7}  \cline{9-10}
H & tx-origin$\triangleleft\ddagger\dagger$ &  & 115 & 477 & M & Uninitialized local$\dagger$ &  & 109 & 824\\ \cline{1-7} \cline{9-10}

H & Integer Overflow$\ddagger\Diamond\bullet\ddagger$ & \multirow{5}{*}{\rot{Arith.}} & 101 & 682 & M & Constant function$\dagger$ &  & - & -\\ \cline{1-2} \cline{4-7} \cline{9-10}
H & Integer Underflow$\ddagger\Diamond\bullet$ &  & 101 & 682 &  M & Shadowing abstract$\dagger$ &  & 119 & 710\\ \cline{1-2} \cline{4-7} \cline{9-10}
M & Signedness bugs$\bullet$ &  & \underline{101} & 682 & M & ERC20 returns false$\triangleleft$ & & \underline{135} & 1164\\ \cline{1-2} \cline{4-7} \cline{9-10}
M & Truncation bugs$\bullet$ &  & \underline{101} & 682 &M & Incorrect Blockhash$\triangleleft$ &  & \underline{104} & 252 \\ \cline{1-7} \cline{9-10}

M & Callstack bug$\bullet\Diamond$ & \multirow{7}{*}{\rot{DoS}} & \underline{113} & 703 &M & Balance Equality$\dagger\triangleleft$ &  &\underline{132} &697\\ \cline{1-2} \cline{4-7} \cline{9-10}
M & Overpowered role$\triangleleft$ &  & - & - &L & Usage of Assembly$\dagger\triangleleft$ &  & - & \underline{695}\\ \cline{1-2} \cline{4-7} \cline{9-10}
M & Gas Limit in Loops$\triangleleft$ &  & \underline{128} & 400 &L & Pragmas version$\triangleleft$ &  & 102 & 937\\ \cline{1-2} \cline{4-7} \cline{9-10}
M & Transfer in Loop$\triangleleft\dagger$ &  & \underline{113} & 703 & L & Should not be view$\triangleleft$ &  & - & - \\ \cline{1-2} \cline{4-7} \cline{9-10}
L & Array Length Manipulation $\triangleleft$ &  & \underline{128} & 400 &L & Bad Visibility$\triangleleft$ &  & 108 & 710 \\ \cline{1-2} \cline{4-7} \cline{9-10}
L & Multiple Calls$\ddagger$ &  & 113 & 703 & L & Shadowing-builtin$\dagger$ &  & 119 & 710  \\ \cline{1-7} \cline{9-10}


H & Reentrancy-eth$\dagger$\# & \multirow{6}{*}{\rot{Reentrancy}} & 107 & 841 & L & Shadowing-local$\dagger$ &  & 119 & 710 \\ \cline{1-2} \cline{4-7} \cline{9-10}
M & Message call to ext. contract$\ddagger$ &  & 107 & 841 &L & Hardcoded address$\triangleleft$ &  & - & \underline{547}  \\ \cline{1-2} \cline{4-7} \cline{9-10}
M & Call without data$\triangleleft$ &  & \underline{107} & 841 & L & Deprecated Constructions$\triangleleft$ &  & 111 & 477 \\ \cline{1-2} \cline{4-7} \cline{9-10}
M & Reentrancy-no-eth$\dagger$\# &  & 107 & 841 & L & Extra gas in loops$\triangleleft$ &  & \underline{128} & 400\\ \cline{1-2} \cline{4-7} \cline{9-10}
L & Reentrancy-benign$\dagger$\# &  & 107 & 841 &L & Redundant fallback reject$\triangleleft$ &  & \underline{135} & 1164\\ \cline{1-2} \cline{4-7} \cline{9-10}
L & State change after ext. call$\ddagger$ &  & 107 & 841 & L & Revert require$\triangleleft$ &  & \underline{123} & 573\\  \cline{1-7} \cline{9-10}

H & Unchecked call return value$\triangleleft\ddagger$ & \multirow{3}{*}{\rot{U.L.C}} & 104 & 252 & L & Exception State$\ddagger$ &  & \underline{110} & 670\\ \cline{1-2} \cline{4-5} \cline{6-10} 
M & Unused return$\dagger$ &  & \underline{135} & 1164 & M & \multicolumn{2}{l|}{Transaction Order Dependence$\ddagger$} &  114 & 362\\ \cline{1-2} \cline{4-10} 
L & Send$\triangleleft$ &  & \underline{104} & 252 & M & \multicolumn{2}{l|}{Timestamp manipulation$\triangleleft\dagger\ddagger\bullet\Diamond$}  & 116 & 829 \\ \hline

\end{tabular}
\begin{tablenotes}
    \item tools: $\dagger$ = Slither, $\ddagger$ = Mythril, $\triangleleft$ = SmartCheck, $\Diamond$ = Oyente, $\bullet$ = Osiris
    \item severity: H = high, M = medium, L = low
    \item \_: inferred by us and not directly present in SWC and CWE, -: not present in the vocabulary
    \item Acc.: Access, Arith.: Arithmetic, U.L.C.: Unchecked Lowlevel Calls
\end{tablenotes}
\end{table*}

Different state-of-the-art approaches use static, dynamic, taint analysis, and symbolic execution of the source code to detect vulnerabilities in SCs. In~\cite{Durieux2020}, the authors analyze the source codes of different SCs for vulnerabilities using nine different SC code analysis tools. Note that we do not survey the different SC code analysis tools as it is out of the scope of this work. However, we provide a brief description and our analysis of results in~\cite{Durieux2020}. As initial results, for the data they had, they found only 42\% of SC had verified unique source codes in which only 4.8\% were unique. The analysis of these unique SCs reveals that the vulnerabilities under the categories such as \textit{Access Control}, \textit{Denial of Service}, and \textit{Front Running}, present under the DASP-10 vocabulary, are not captured well by most of the tools.
Further, they found that analysis tools such as Mythril~\cite{mueller2018smashing} and Slither~\cite{Feist2019} together identify the maximum vulnerabilities present in the DASP vulnerability set. Individually, Mythril performs the best and identifies 27\% of vulnerabilities present in their dataset. Note that slither uses a call graph to identify vulnerabilities. In another work, in~\cite{bis2020}, the authors compare different SC analysis tools such as Remix, Slither, SmartCheck, Oyente, Mythril, and Securify and find out that Slither performs the best as it detects at least one vulnerability from different vulnerability classes considered by them. In~\cite{Angelo2019}, the authors categorized the SC vulnerabilities into three groups: blockchain platform-based vulnerabilities (such as \textit{Transaction Ordering Dependence (TOD)}, \textit{Random Number}, \textit{Timestamp}), EVM based (such as \textit{CallStack Depth}, \textit{Lost Ether}), and Solidity based (12 vulnerabilities including \textit{Reentrancy}, \textit{Unchecked Calls}, and \textit{tx.origin}). They analyze these vulnerabilities using 27 different SC analysis tools. They concluded that Mythril could identify 75\% of the blockchain-based vulnerabilities, while SmartCheck~\cite{Tikhomirov2018} detects 72\% of all the vulnerabilities. In~\cite{parizi2018}, the authors also reach a similar conclusion. 

In~\cite{Wang2020}, the authors study six types of vulnerabilities: \textit{Integer Overflow and Underflow}, \textit{Transaction-Ordering Dependence}, \textit{Callstack Depth Attack}, \textit{Timestamp Dependency}, and \textit{Reentrancy Vulnerabilities} while not considering \textit{Denial of Service (DoS)} and \textit{tx-origin}. They develop an ML-based model called \textit{ContractWard}, which at first uses `SMOTE' for over-sampling data and then under-sampling data points that have neighborhood relations. They then extract 1619 features using 2-gram analysis of opcodes of SCs to automatically detect the vulnerabilities mentioned above and then apply `XGBoost'. Here, 2-gram refers to a set of 2 tokens, where the probability of occurrence of a token depends on the previous token.  They achieve an average F1-score of 0.96 on their dataset. However, ContractWard has two limitations: \textit{(i)} it uses `SMOTE' to oversample dataset, and \textit{(ii)} do not consider transactions carried out by SCs, which are important as not all vulnerable SCs are necessarily exploited~\cite{Perez2021}. 

\subsection{Transaction Based Techniques}\label{sec:ML}

In~\cite{agarwal2020detecting}, the authors use the transactions of accounts on the Ethereum blockchain to develop temporal transaction features to identify malicious accounts. They first survey different state-of-the-art algorithms used to detect malicious accounts on different permission-less blockchains. They identify that transactions on the Ethereum blockchain show bursty behavior for the degree, gas-price, inter-event time, and balance. Using burst-based features, the authors developed an ML pipeline that achieves high recall ($>78\%$) in detecting the entire malicious class in their dataset. However, they considered SCs as equivalent to EOAs and did not consider internal transactions. In a follow-up study, in~\cite{agarwal2021detecting}, the authors analyze different malicious activities and identify that neural networks perform best while detecting any adversarial attack that uses transaction behaviors as a feature vector.

In~\cite{Hanyi2019}, the authors cluster EOAs and SCs in the Ethereum  based on their transactions. They use a dataset containing transactions of 526121 accounts. They use the \textit{birch} algorithm to perform hierarchical clustering on their dataset and only study the top 10 clusters with the maximum number of accounts. They observed that many malicious accounts cluster together. 

In another approach, in~\cite{ChenL2021}, the authors use Graph Convolutional Networks (GCN) to detect EOA and SCs associated with Phishing activities. They first acquire transactions of accounts marked as `Fake Phishing' from Etherscan and build a graph that has accounts as its nodes and the transactions carried out by them as edges. They obtain a graph with 13 connected components where they choose only the largest connected component subgraph for their analysis. They use features such as `Indegree', `Outdegree', and `Number of Neighbors' to detect Phishing accounts. On the feature vector, GCN achieves an average F1 score of 0.24. Although they consider both internal and external transactions, they assume that all the accounts which are related to or carry out internal transactions are `non-phishing' EOAs and SCs.

In~\cite{FARRUGIA2020}, the authors use XGBoost to detect illicit EOAs and SCs (including tokens such as ERC-20) in the Ethereum blockchain. They use only 2179 malicious accounts and 2502 normal accounts for their experiments and use 42 transaction-based features such as \textit{Time\_diff\_between\_first\_and\_last}, \textit{min\_value\_received} and \textit{min\_value\_sent}. They also rank features based on their importance and find \textit{Time\_diff\_between\_first\_and\_last} as the most important feature while \textit{ERC20\_Most\_Sent\_Token\_Type} being the least important one. Using these features, they achieve an average accuracy of 0.963. However, they use a highly under-sampled dataset, with a ratio of 1:1.14 between the malicious and the benign classes. This does not represent the actual distribution of malicious and benign accounts in Ethereum, which has more than 14 million unique accounts.

Note that the above state-of-the-art approaches do not consider source code-based features to carry out a behavioral analysis on the SC in Ethereum. To address such an issue, in~\cite{Camino2020}, the authors use both transaction and source code-based features to detect honeypot accounts in the Ethereum blockchain. They use a dataset with 16163 accounts, of which 295 are marked as honeypots by HoneyBadger's repository~\footnote{\url{https://github.com/christoftorres/HoneyBadger}}. Using transaction-based features (such as \textit{Transaction Count} and \textit{Transaction Value}) and source code-based features (such as \textit{hasByteCode} and \textit{hasSourceCode}), the authors train `XGBoost'. They achieve an `Area Under the Receiver Operating Characteristics' (AUROC) mean score of 0.968 on their dataset. They, however, do not consider the temporal aspects of blockchain transactions. 

Moreover, in~\cite{Camino2020}, the authors use features extracted from the opcodes of an SC and their transactions. However, they do not consider the vulnerabilities that are present and are exploited by attackers as features. To the best of our knowledge, ours is the first work in the field of blockchain security that considers both temporal behavior (extracted using both internal and external transactions) and vulnerabilities present in an SC to detect potential suspects. 

\section{Methodology}\label{sec:method}

In this section, we describe our approach in detail. We first obtain the source codes of all the SCs available in the Ethereum blockchain using the Etherscan APIs~\cite{etherscanApi} and their malicious tags (using the Etherscan label cloud service~\cite{etherscanLC}) and the internal and external transaction data.

\begin{figure}
    \includegraphics[width=0.5\textwidth]{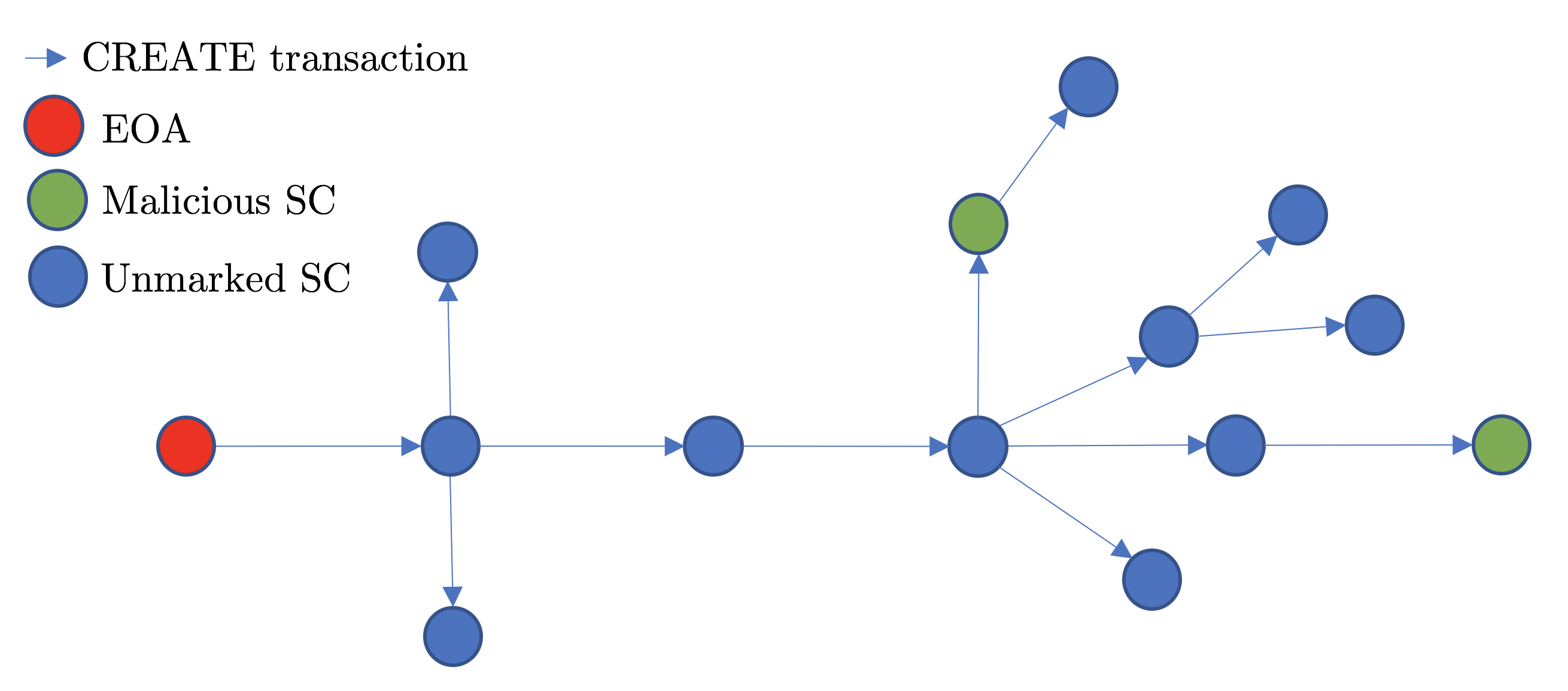}
    \caption{Sample graph component constructed using CREATE transactions of malicious SCs, their successors, and their predecessors.}
    \label{fig:graph}
\end{figure}

With CREATE-type internal transactions, a malicious SC can create several child SCs. These child SCs further develop several other SCs. Most child SCs have the same source code as their malicious parent SC; therefore, their vulnerabilities are also the same. In general, despite not showing malicious transaction behavior, such child SCs could also be exploited and thus should be marked as suspects. On the other hand, a parent of the malicious SC could be unaware of the vulnerabilities or could have written the malicious SC with specific malicious intent. Therefore, parents of malicious SC should also be marked as suspects. This leads to marking the entire parent-child chain of the malicious SCs as malicious. Contrary to this, in specific cases, valid organizations create numerous user-centric SCs where only a particular SC is involved in performing a malicious activity (for example, Bittrex). In such cases, marking the entire chain suspicious would be incorrect. Thus, where we know that an organization developed a particular SC for a specific purpose, we do not create its chain. While for other malicious SCs we do. Figure~\ref{fig:graph} depicts a sample graph component generated using CREATE transactions of malicious SCs, their successors, and their predecessors. Currently, in state-of-the-art approaches, such SCs are not considered while training ML algorithms. We thus include such SCs as malicious SCs in our study.

Next, we identify code similarity between all the suspect SC identified using the above assumption. Here, we use the hashing technique as used in~\cite{Durieux2020}, where we generate a hash of the source code and identify unique hashes. Note that if two or more SCs have different source codes, they are lexicographically different, resulting in SCs having different hashes. We do not consider OPCODEs to generate the hashes. Hashes reduce the computational resources needed to identify vulnerabilities in the SCs as SCs with identical hashes will have the same vulnerabilities. We analyze the uniquely identified SCs with five different vulnerability detection tools: SmartCheck, Mythril, Oyente, Slither, and Osiris. While Slither and SmartCheck carry out static analysis on SC source code, Mythril relies on taint analysis. Oyente uses symbolic execution of the SC source code to detect vulnerabilities, while Osiris extends this functionality to detect Integer related bugs. Here, we choose these tools as they collectively detect most of the vulnerabilities described in Table~\ref{table:vulnerabilityMapping}.

Along with the ground truth available about the type of malicious activity a particular SC is involved in and all the identified SC vulnerabilities, we then study the correlation between the vulnerabilities and the type of malicious activity they are associate with. In the process, we also study if exploitability is reflected due to the existence of any vulnerability. A severity score is associated with all the vulnerabilities present in the SCs. Also, an SC can have multiple vulnerabilities with different severity. Let $V^i$ be the set of all vulnerabilities present in an SC $i$ and let a vulnerability $j\in V^i$ have a severity score $S_j$. A \textit{severity score} ($Ss_i$) for an SC $i$ is defined as equation~\ref{eq:sevScore}. Note that the $Ss^i$ represents the average severity score.

\begin{equation}\label{eq:sevScore}
    Ss^i = \frac{\sum_{\forall j \in V^i} S_j}{||V^i||}
\end{equation}

Besides computing the severity score, we analyze the transaction dataset and identify different behavioral features. These features are based on the approach defined in~\cite{agarwal2020detecting} (these features are also listed in Appendix A2). These features capture \textit{(i)} temporal behavior along with static properties, and \textit{(ii)} provide the best results in terms of recall on malicious class. For our study, we create two different data configurations using such features (one with the severity score defined above and another without). In this work, we study different unsupervised ML algorithms such as K-Means, HDBSCAN, Spectral, Agglomerative clustering, and OneClassSVM to identify the algorithm (and their hyperparameters) that perform the best based on silhouette score Table~\ref{table:Hyperparams} lists different hyperparameters that we test for a particular algorithm. The choice of hyperparameter values reflects the computational power available to us. Ideally, to establish correlation, SC's should cluster better when we use the severity score as a feature along with the transaction and temporal-based features. 

\begin{table}
\caption{Algorithms and Hyperparameters Tested}\label{table:Hyperparams}
\begin{tabular}{|l|c|}
\hline
Algorithm & Tested on Hyperparameters\\ \hline
K-Means  & n\_clusters $\in[3,26]$ \\ \hline
HDBSCAN & min\_cluster\_size $\in[2,1735]$\\ \hline
Spectral Clustering & n\_clusters $\in[3,26]$\\ \hline
Agglomerative Clustering & n\_clusters $\in[3,26]$ \\ \hline
OneclassSVM & Degree $\in[2,10]$, kernel=`poly'\\ \hline
\end{tabular}
\end{table}

We analyze the datasets based on different temporal granularities to understand behavioral aspects and whether SCs show persistent malicious behavior over time. In a particular temporal granularity, we consider only those SCs and their transactions that occur in a specific period defined by the temporal granularity. For instance, in a 1-Day temporal granularity, we consider only the transactions in a given 1-Day period. Note that there could be multiple periods in a particular granularity. We study the behavior across four different temporal granularities in this work: 1-Day, 3-Day, 1-Month, and aggregated (ALL). Henceforth, whenever we use the word `segment', we refer to a particular period from a granularity as mentioned above. We create two different feature vectors for each segment: one that contains both severity score and transaction-based features and another that only has transaction-based features. For each segment and the two sets of feature vectors, we use the unsupervised algorithm that performs best (identified via Q2) with an assumption that the same algorithm would serve the best across the different segments. For each segment in each granularity, we then determine the largest cluster with a maximum number of malicious SCs. We then compute the cosine similarity amongst benign and malicious SCs present in that cluster to identify which benign/unmarked SCs behave similarly to the malicious SCs. Our motivation for using the largest cluster only is: \textit{(a)} we assume that all malicious SCs show similarity and cluster together, and \textit{(b)} choosing such cluster reduces the search space and is within the computation limits available to us. We acknowledge that there exist several other metrics to identify similarity scores (such as \textit{Jaccard}), and using such metrics will give different results. But we use the cosine similarity metric because \textit{(i)} it is more popular and widely adopted, and \textit{(ii)} was used in~\cite{agarwal2020detecting} (we use their features) to detect the suspects. A very high cosine similarity score ($CS_{ij}\rightarrow1$) between a malicious SC, $i$, and a benign SC, $j$, indicates that $j$'s behavior is suspicious in that segment. For each granularity, as $j$ could change its behavior over time, we associate $j$ with a probability ($p(j)$) of being malicious. Over all the segments, a high probability ($p(j)=1$) means that the SC should be marked malicious considering that temporal granularity. To identify this probability, we use the same method as in~\cite{agarwal2020detecting}. We then compare across different granularities to determine which suspect SCs are common to say that the used granularity does not impact their behavior.

\begin{figure*}
    \includegraphics[width=0.7\textwidth]{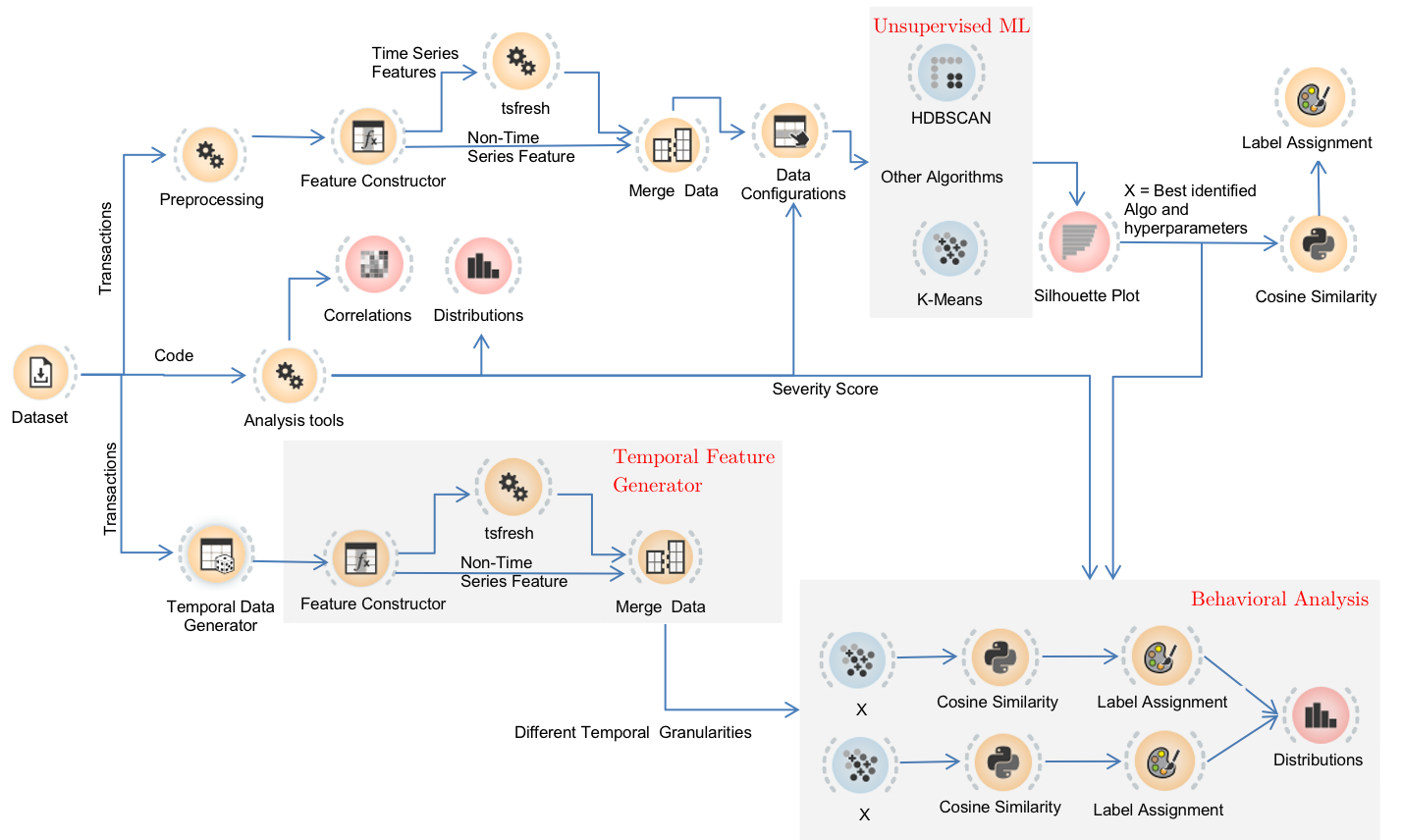}
    \caption{ML Pipeline}
    \label{fig:mlp}
\end{figure*}

In summary, Figure~\ref{fig:mlp} presents the entire pipeline. We first use SC analysis tools to do a code-based analysis and detect vulnerabilities in the source code of SCs present in our dataset. We then correlate the detected vulnerabilities present in SCs with different malicious activities. Since each vulnerability has an associated severity, we assign a severity score to each SC. This score is averaged over all the vulnerabilities present in that SC. To determine the usefulness of the severity score towards detecting malicious SCs, we first compute different transaction and graph-based temporal features. We then create different datasets, one with both transaction and severity-based features and the other with only transaction-based features. We then use different unsupervised ML algorithms on the above-created datasets across different temporal granularities (such as 1-Day, 3-Day, and 1-Month) and analyze the results to determine the usefulness of the severity score as a feature.

\section{Evaluation and Results}\label{sec:eval}

This section provides an in-depth analysis of our approach towards answering Q1, Q2, and Q3 and presents our results. All our analysis is performed using Python version 3 and its associated libraries such as scikit-learn.

\subsection{Data}\label{sec:data}

We use the Etherscan blockchain explorer APIs~\cite{etherscanLC} to acquire SCs which are associated with malicious activities. This results in a list of 403 SCs marked malicious until 28th August 2020 (block number 10747845) since the induction of Ethereum. Note that the tag (including malicious ones) assigned to an SC in Etherscan is crowd-sourced, i.e., any person can suggest the tag. Since we cannot validate the correctness of these tags, we assume that SCs are correctly associated with different malicious activities. Although there are multiple malicious tags present via Etherscan, such as those described in~\cite{agarwal2021detecting}, until the data collection time, malicious SCs are only associated with four malicious activities: Phishing, Gambling, High-Risk, and Ponzi Scheme.

For all these 403 SCs, using their internal transactions and the heuristics described in Section~\ref{sec:method}, we identify~$\approx2$ Million SCs that are either successor or predecessor to a malicious SC. For all the marked malicious SCs, we observe that:
\begin{itemize} 
    \item 377 out of 403 marked SCs, are created by EOAs. These SCs do not have any CREATE type transactions and thus do not create any successor SCs.
    \item Out of the remaining 26 marked malicious SCs, only 8 SCs (that also have EOAs as their parents) create a total of 52 SCs. However, these 52 SCs did not create any new SCs.
    \item Out of the remaining 18 marked malicious SCs, these SCs have 12 unique SCs as their parent. Although these 18 SCs did not create any successor SCs, their 12 parents created many SCs.
\end{itemize}

Out of all the SC, we observe that only 46 unique hashes exist and corresponding to 46 unique SC codes. Note that this number represents SCs for which source codes are available. There are 165 malicious SCs for which the source code is not available. We do not consider them in our study because the source code could be different and our feature vector depends on the severity score obtained using vulnerabilities present in the SCs. In all 46 unique hashes identified, 38 unique hashes belong to 38 marked SCs. This also means there are only 38 unique codes present between the remaining 238 marked SCs. In the remaining 8 SCs that are unmarked and detected using our graph analysis, 7 SCs lie in the graph's component created using seven different marked Phishing SCs, and 1 SC lies in the graph component created using 1 marked Ponzi scheme-based SC. As we have limited computational resources, analyzing both internal and external transactions of all these 2 million SCs is practically not feasible for us. Thus, we restrict our analysis and consider a union of these 46 SCs and 47398 unique SCs identified by~\cite{Durieux2020}.

We identify 314614 vulnerabilities in total. Out of these, 314302 are present in the benign SCs, and 312 are present in malicious SCs across different severities (54 for high, 92 for medium, and 166 for low). These 312 vulnerabilities in the 46 malicious SCs are distributed as follows: 192 vulnerabilities are present in the Phishing based SCs, 95 vulnerabilities are present in the Gambling based SCs, 19 vulnerabilities are present in the High-Risk based SCs, and 6 vulnerabilities are present in the Ponzi scheme-based SCs. In Phishing SCs, there are 26 vulnerabilities with high severity, 62 with medium, and 104 with low severity. Gambling SCs have 21 vulnerabilities with high severity, 20 with medium, and 54 with low severity. High-Risk SCs have 6 vulnerabilities with high severity, 7 with medium, and 6 with low severity. Similarly, Ponzi scheme-based SCs have 1 vulnerability with high severity, 3 with medium, and 2 with low severity. 

\begin{figure*}
    \centering
    \subfloat[in Month Granularity][in Month Granularity]{
        \includegraphics[width=0.5\textwidth]{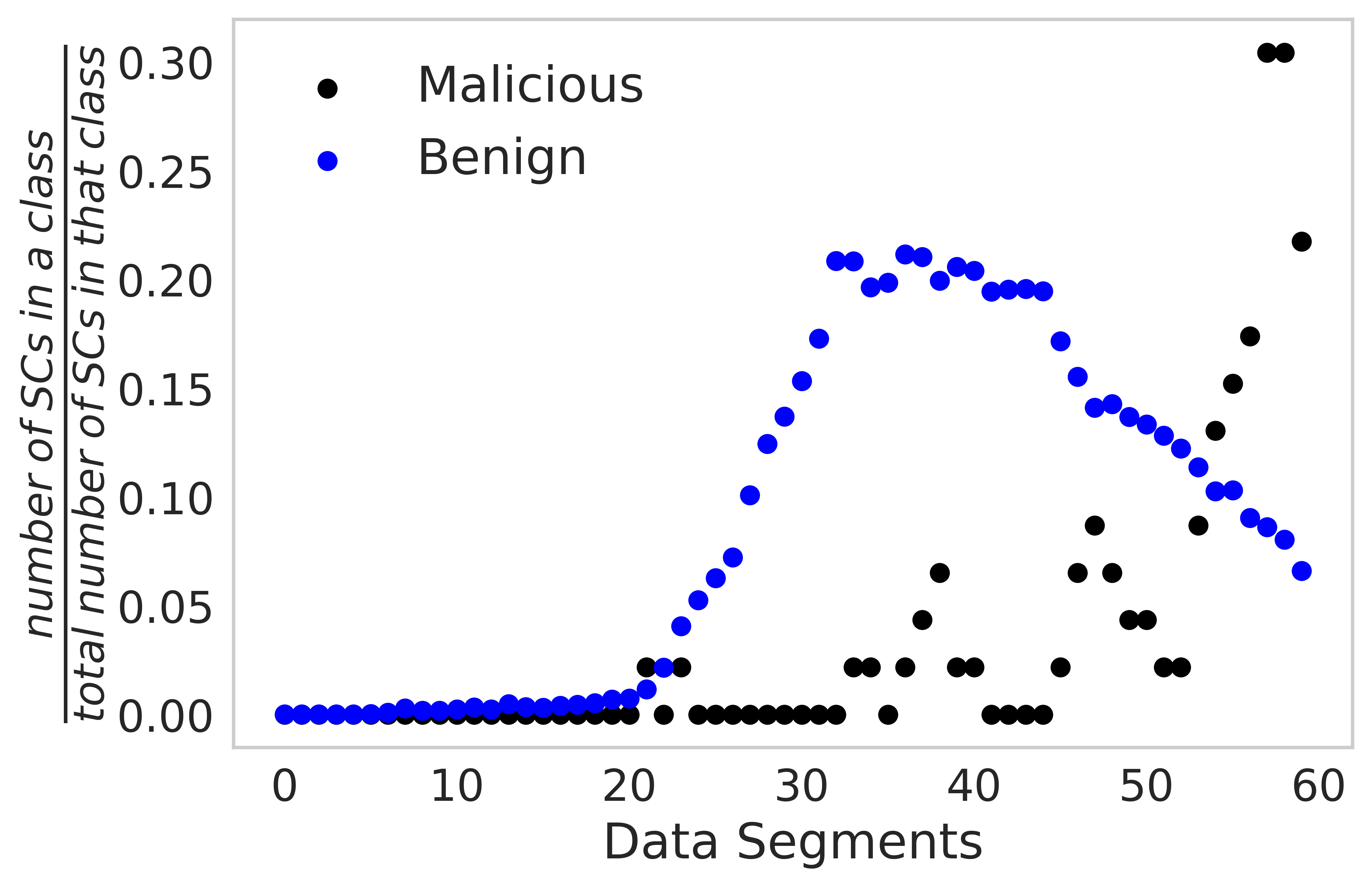}
        \label{fig:monthdistro}
    }
    \subfloat[in 3-Day Granularity][in 3-Day Granularity]{
        \includegraphics[width=0.5\textwidth]{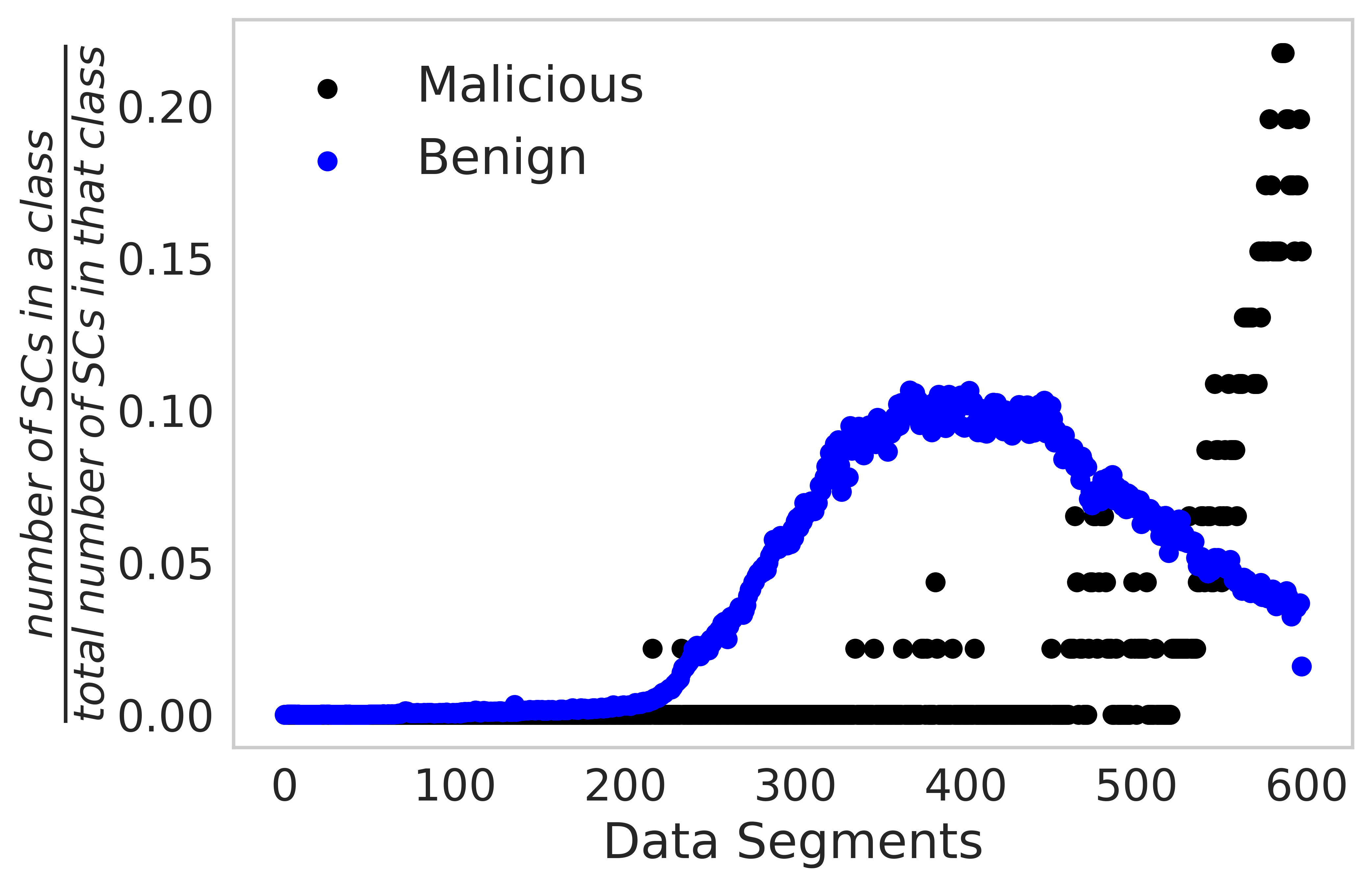}
        \label{fig:3daydistro}
    }\\
    \subfloat[in Day Granularity][in Day Granularity]{
        \includegraphics[width=0.5\textwidth]{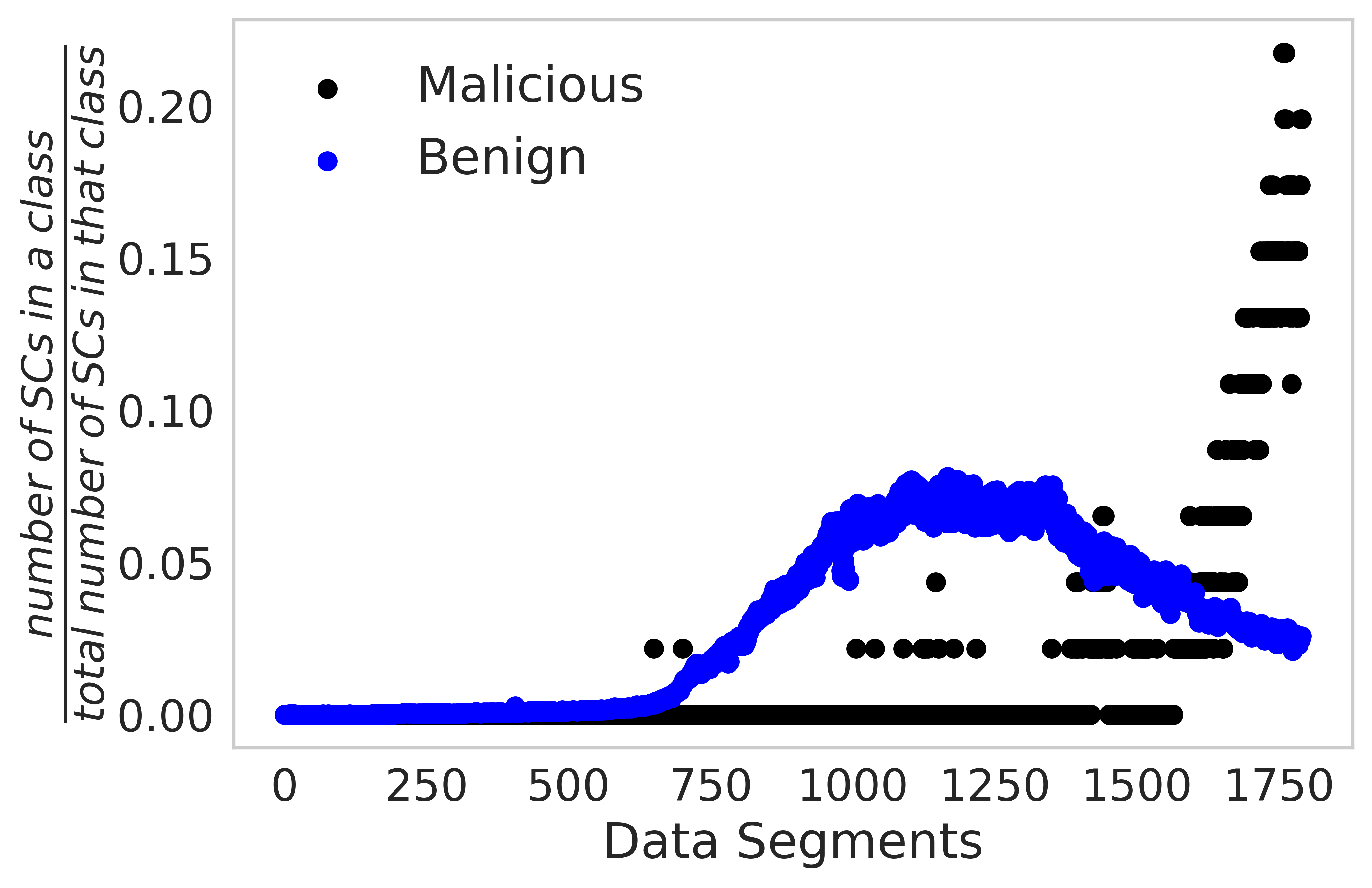}
        \label{fig:daydistro}
    }
    \caption{Fraction of the number of SCs of particular class to the total number of SCs of that class in a temporal granularity}\label{fig:temporaldistribution}
\end{figure*}

In Ethereum, on average, their are~$\approx6000$ blocks created each day. Using such information, we develop segments for different temporal granularities. For 1-Day granularity, from the genesis block until our collection date, we have 1791 segments. Here each segment corresponds to 6000 blocks. For example, segment 1 contains transactions of considered SCs from genesis block until block number 6000. Similarly, for the segments in the 3-Day granularity, we consider transactions of considered SCs in 6000$\times$3 blocks, and for segments in 1 month, we consider 6000$\times$30 blocks. Thus we have 598 segments and 60 segments for the 3-Day granularity and the Month granularity, respectively. Figure~\ref{fig:temporaldistribution} shows the distribution of the fraction of both benign and malicious SCs that transacted in each of the above-mentioned temporal granularities over the total benign and malicious accounts considered. Here, we observe that the fraction of malicious SCs increases across different temporal granularities representing that malicious activity has increased with time and adoption of Ethereum.

\subsection{Results}\label{sec:results}

Our results pertain to the three research questions. Thus this section is divided into 3 parts, with each part referring to the research question defined in Section~\nameref{sec:intro}.

\begin{figure*}
    \centering
    \subfloat[Count of malicious SCs with a particular vulnerability in a specific type of malicious activity][Count of malicious SCs with a particular vulnerability in a specific type of malicious activity]{
        \includegraphics[width=0.5\textwidth]{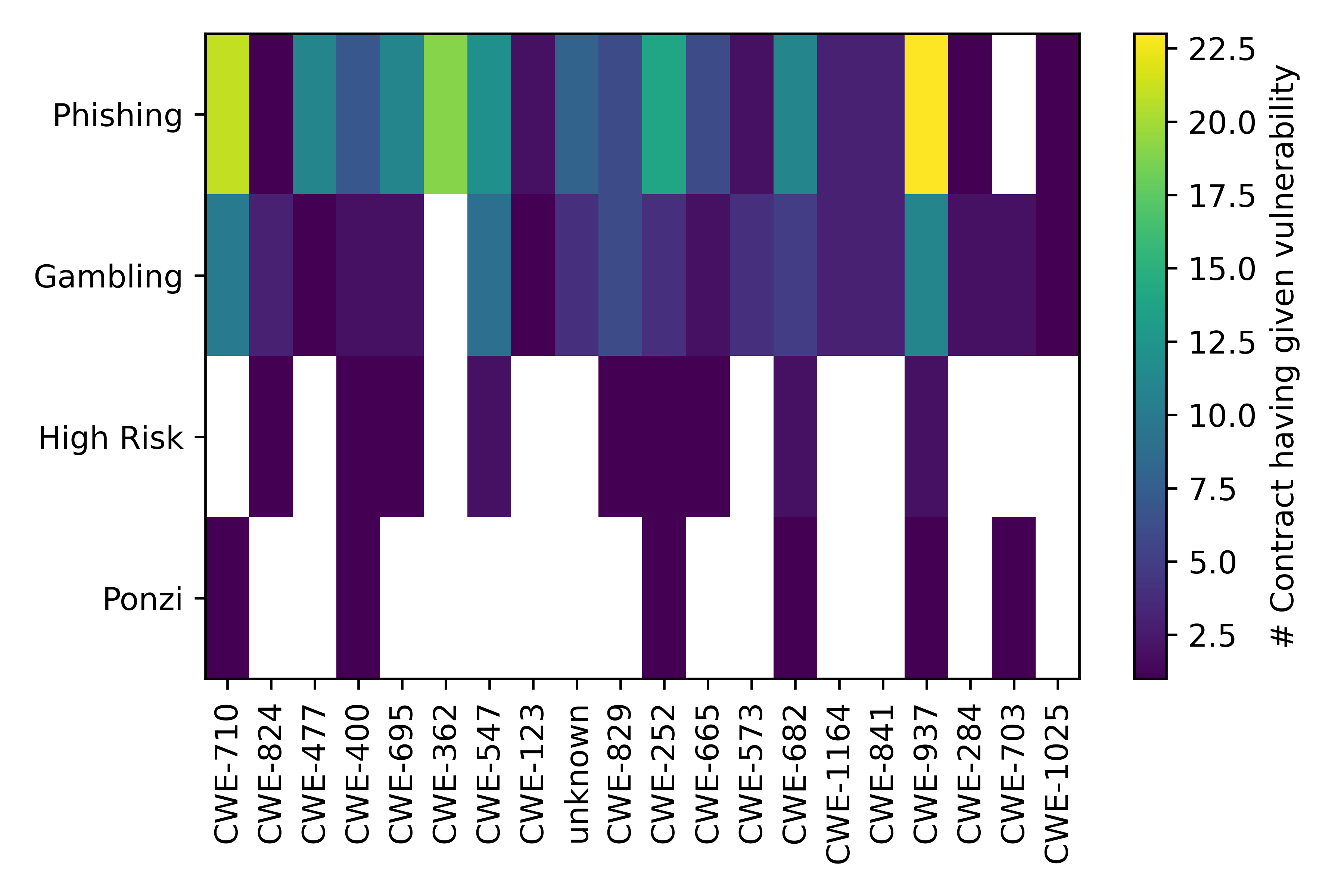}
        \label{fig:vulnDistroMal}
    }
    \subfloat[Count of malicious SCs with a particular vulnerability in a specific type of malicious activity normalized by the number of SCs of that malicious type. The number in ``()'' on the y-axis represents the number of SC of particular malicious activity][Count of malicious SCs with a particular vulnerability in a specific type of malicious activity normalized by the number of SCs of that malicious type. The number in ``()'' on the y-axis represents the number of SC of particular malicious activity]{
        \includegraphics[width=0.5\textwidth]{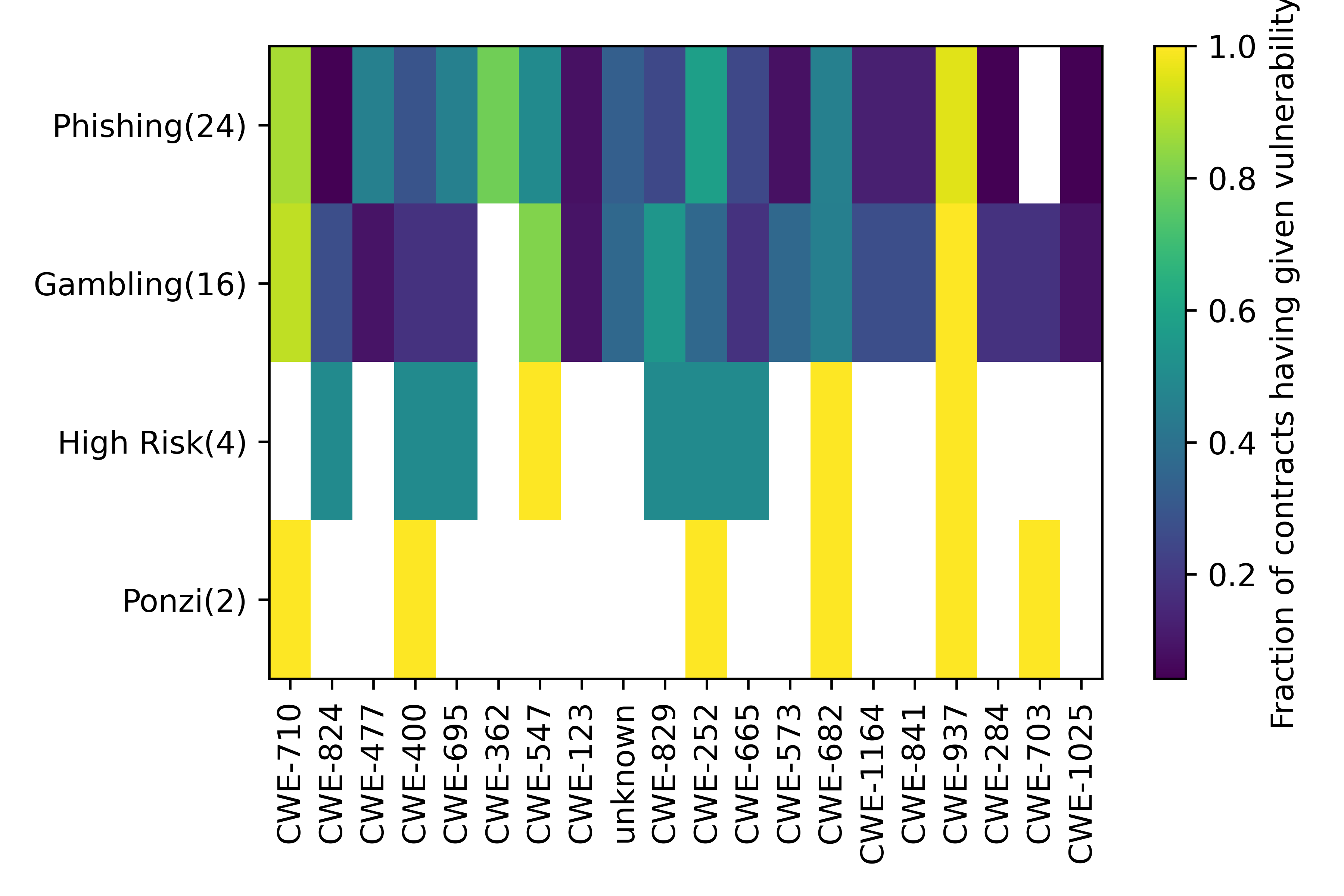}
        \label{fig:vulnDistroFractionMal}
    }
    \caption{Distribution of vulnerabilities in every type of malicious activities present in the dataset.}\label{fig:vuln}
\end{figure*}

\begin{figure*}
    \centering
    \subfloat[Number of vulnerabilities of specific type across all SCs on a semi-log scale][Number of vulnerabilities of specific type across all SCs on a semi-log scale]{
        \includegraphics[width=0.5\textwidth]{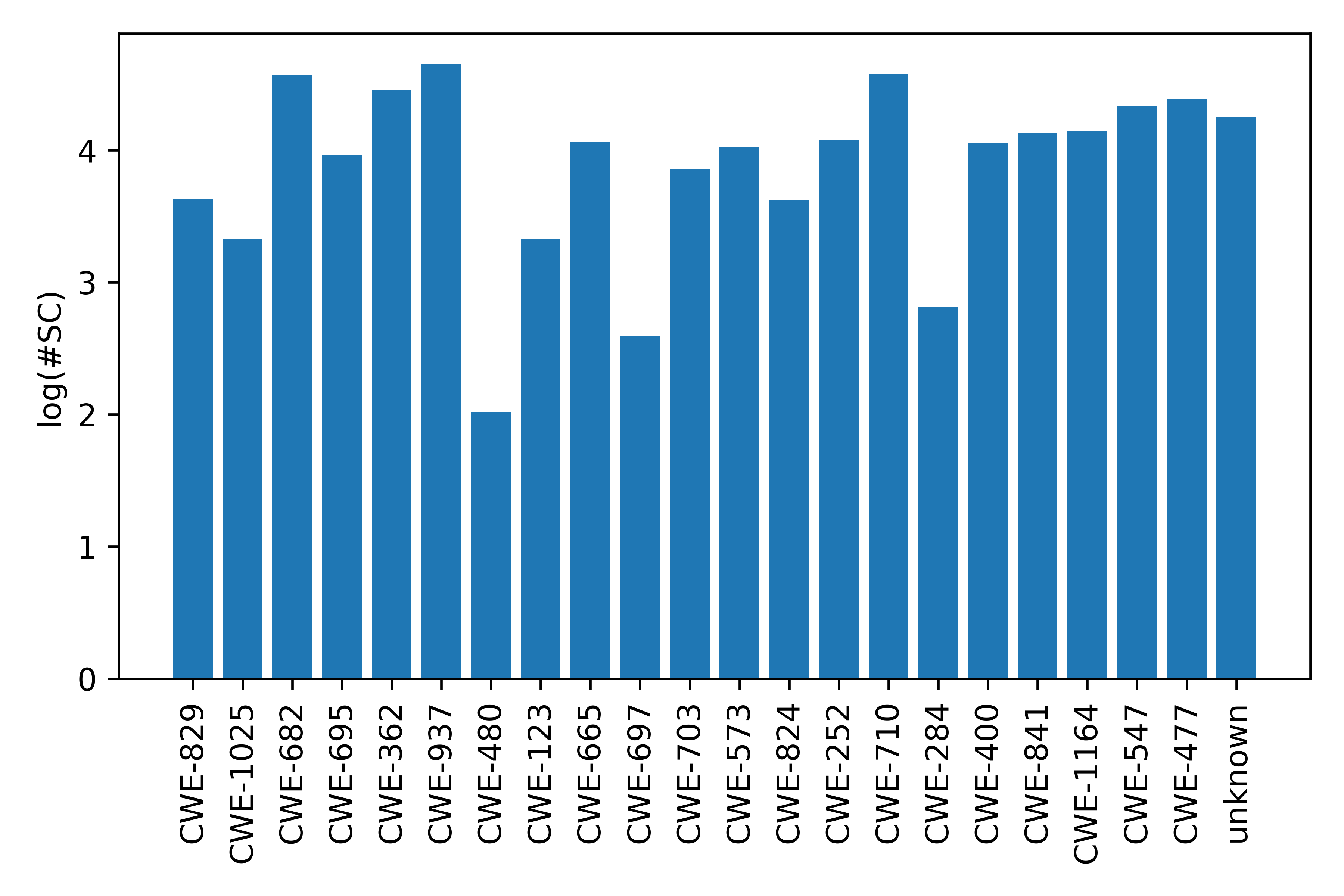}
        \label{fig:vulnDistro}
    }
    \subfloat[Fraction of vulnerabilities of a particular severity in SCs][Fraction of vulnerabilities of a particular severity in SCs]{
        \includegraphics[width=0.5\textwidth]{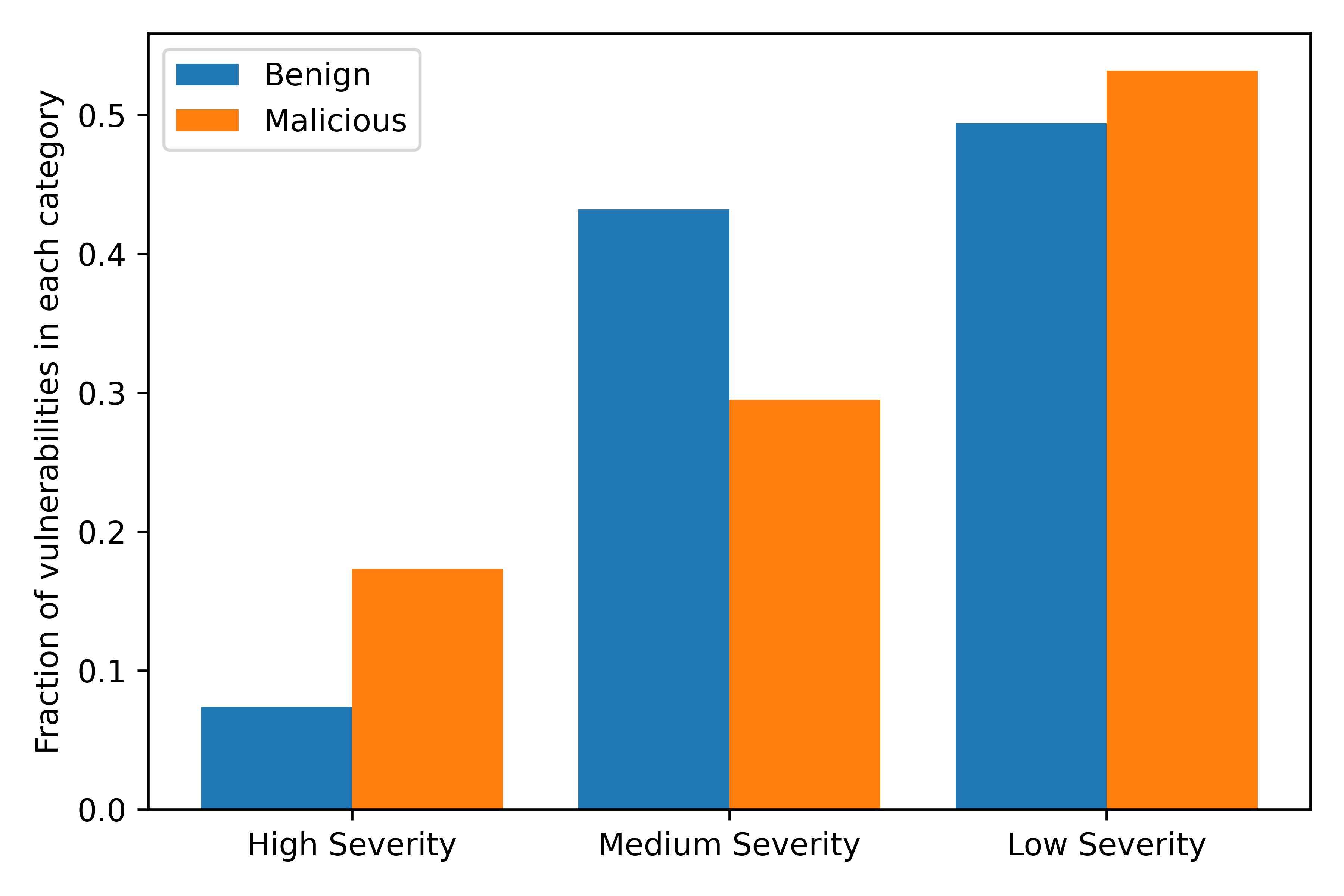}
        \label{fig:severityDistro}
    }
    \caption{Distribution of Vulnerabilities}\label{fig:distribution}
\end{figure*}

\subsubsection{Q1: Correlation between malicious activities and Vulnerabilities and whether severity of a vulnerability correspond to its exploitability}

One way to identify the correlation between malicious activities and vulnerability is to study the distribution of the vulnerabilities in the malicious activities associated with the SCs. Thus, we identify the fraction of malicious contracts related to a specific CWE vulnerability for each category of malicious activity. To understand the correlation, we normalize this and then study the relation, if any. Figure~\ref{fig:vuln} shows both the number (cf. Figure~\ref{fig:vulnDistroMal}) and the normalized count of a specific vulnerability (cf. Figure~\ref{fig:vulnDistroFractionMal}).
From Figure~\ref{fig:vuln}, we identify that none of the SCs related to Phishing type of malicious activity have CWE-703 vulnerability. However, vulnerabilities such as CWE-362, CWE-937, CWE-252, and CWE-710 are present in large numbers, with CWE-937 present in almost all malicious SCs. Here, we also note that CWE-362 (a medium severity vulnerability) is only present in the SCs related to the Phishing type malicious activity. Moreover, SCs tagged under malicious activities such as High-Risk, Gambling, and Ponzi labels do not report SC vulnerabilities under CWE-362. 
Further, we observe that SCs under the Ponzi scheme have 6 vulnerabilities: CWE-710, CWE-400, CWE-252, CWE-682, CWE-937, and CWE-703. 
We also observe that although each malicious activity has SCs with high severity vulnerabilities, such as CWE-841 and CWE-123, their frequency is less. From the above observations, we infer that in SCs corresponding to malicious activities, the vulnerabilities with high severity are less in number. However, just from the above observations we cannot say that, for example, if an SC has vulnerabilities related to CWE-362, it is involved in Phishing activity. Our inference is based on the fact that CWE-362 is also present in benign SCs. Nonetheless, it is possible that such benign SCs are, in reality, related to Phishing activity but are not marked as Phishing SCs. 

Further, to check if benign SCs also have vulnerabilities, we plot the distribution of vulnerabilities. Figure~\ref{fig:vulnDistro} shows the distribution of the vulnerabilities across all the SCs on a semi-log scale. Here, we observe that vulnerability CWE-937 is most frequent and occurs in most SCs (including benign SCs), while vulnerability CWE-480 is least common. With respect to the severity score (cf. Figure~\ref{fig:severityDistro}), we observe that high severity vulnerabilities (23214 vulnerabilities in total) are also present in the benign SCs, but their fraction (= (number of high severity)/(total vulnerabilities in the considered class)) is less than those present in the malicious SCs. We observe similar behavior for low severity vulnerabilities. However, in this case, the fraction is much higher. On the other hand, the fraction is much higher for the benign class for the medium severity vulnerabilities. Upon further investigation, we find a negligible difference between the average severity score of benign SCs (= 2.25) and malicious SCs (= 2.21). Further, as the difference between the fraction for malicious and benign class for each severity category is very small, we cannot say that severity of a vulnerability relates to exploitation.


\subsubsection{Q2: Importance of Severity score}

To identify the importance of the severity score, we test the results obtained using different unsupervised ML algorithms and different data configurations. We find that K-Means performs best as it achieves the highest silhouette score amongst the various unsupervised algorithms listed in Table~\ref{table:Hyperparams}. For K-Means, when using both transaction and severity score-based features, for $K=18$, the best silhouette score ($0.26$, while for other values of $K$, the silhouette score $\in [0.23, 0.26]$) is achieved. Similarly, when considering only transaction-based features, for $K=25$, the best silhouette score ($0.29$, while for other $K$ silhouette score $\in [0.24, 0.29]$) is achieved. Other unsupervised ML algorithms achieve lesser silhouette scores than K-Means. For HDBSCAN and the tested hyperparameter configurations, we obtain silhouette scores $\in [-0.11, 0.14]$. Here, for most of the hyperparameter configurations, the silhouette score was negative. For Spectral Clustering, the silhouette scores obtained were $\in [0.15, 0.22]$. While for the Agglomerative Clustering, we obtain silhouette scores $\in [0.12, 0.23]$. Finally for the OneClassSVM, silhouette score ranges between $[0.02, 0.09]$. From this analysis, we observe that K-Means provides better silhouette scores. Therefore, we use K-Means for our further analysis. 

\begin{figure}
    \includegraphics[width=0.5\textwidth]{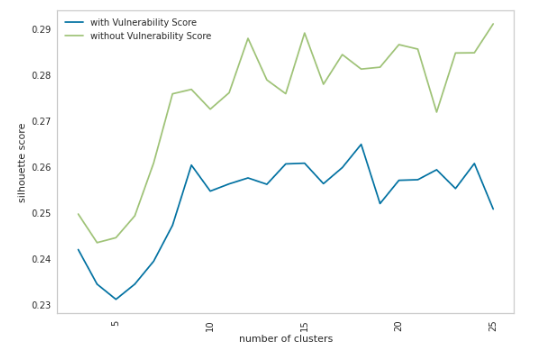}
    \caption{Comparison between Silhouette scores.}
    \label{fig:silheoutteComparison}
\end{figure}

In the case of K-Means, Figure~\ref{fig:silheoutteComparison} shows the silhouette scores for different $K$ for the two different data configurations. From the figure, we observe that the silhouette scores obtained using severity score and transaction-based features always remain less than the scores obtained on using only transaction-based features. Thus, we infer that when such severity scores are considered a feature vector, the data is either more uniformly distributed or more densely distributed in a small feature space, causing overlapping clusters. The clusters thus formed are indistinguishable from each other, which in turn reduces the silhouette score. This also means severity score for inter-cluster analysis is not a good feature. However, severity score for intra-cluster analysis is a good feature to include as it is able to detect more SCs that show high similarity along the feature space with the malicious SCs. 

Nonetheless, we calculate the similarity amongst benign and malicious SCs in the cluster with maximum malicious SCs. We find that the maximum similarity score between malicious and benign SCs is 0.74 when we consider both severity score and transaction-based features. This similarity score reduces to 0.73 when we consider only transaction-based features. The difference in these scores indicates that upon considering severity scores as features, the SCs have less distance between them in the feature space, i.e., are more closer. 
As the maximum is $\in \{0.73, 0.74\}$, we consider this to identify the benign SCs that are suspects and are within $\epsilon=10^{-7}$. We find that there are 2 such benign SCs in both the cases. To further analyze the behavioral changes in the SCs, we identify the probability of an SC being a suspect. Note that in the ALL granularity, probability computation does not make sense as there is no notion of behavior change.

\subsubsection{Q3: Understanding behavioral changes over time}

Using the best-unsupervised ML algorithm identified (K-Means algorithm) and the different data segments created using different temporal granularities, we investigate the cluster with the maximum number of malicious SCs. For temporal granularities other than the ALL granularity, we calculate the probabilities of benign SCs being malicious. Towards this, we run the K-Means clustering algorithm across all the temporal granularities and select the $K$ (number of clusters) for which the maximum silhouette score is obtained for our analysis. Again, we investigate the cluster where the maximum number of malicious SCs are present for different temporal granularities. From the selected cluster, we then select those benign SCs as suspects where the $CS\rightarrow1$ with malicious SCs, i.e., lie $\in 1-\epsilon$ where $\epsilon=10^{-7}$. Here, we find that:
\begin{itemize}
    \item 1066, 24, and 4 SCs are identified as suspects and have $p=1$ in 1-Month, 3-Day, and 1-Day granularity, respectively, when both transaction and Severity scores are used as features. Here, we do not identify any suspect SC that appeared across different temporal granularities. 
    \item 866, 24, and 2 SCs are identified as suspects and have $p=1$ in 1-Month, 3-Day, and 1-Day granularity, respectively, when only transaction-based features are used. Here as well, we do not find any suspect SC which is common across different temporal granularities.
    \item In these identified suspect SCs for the above two cases (when using severity score along with temporal features and when only using temporal features), we again do not find any common suspect SC. 

\end{itemize}

Note that the difference in numbers of SCs identified as malicious for the two cases (with and without using severity score) is due to the reasons described in previous sub-section. That is, the data points become well-clustered when both severity score and transaction-based features are used, thereby increasing the intra-cluster density where the similarity score between malicious and benign SCs is high.

\begin{figure*}
    \centering
    \subfloat[in Month Granularity][in Month Granularity]{
        \includegraphics[width=0.5\textwidth]{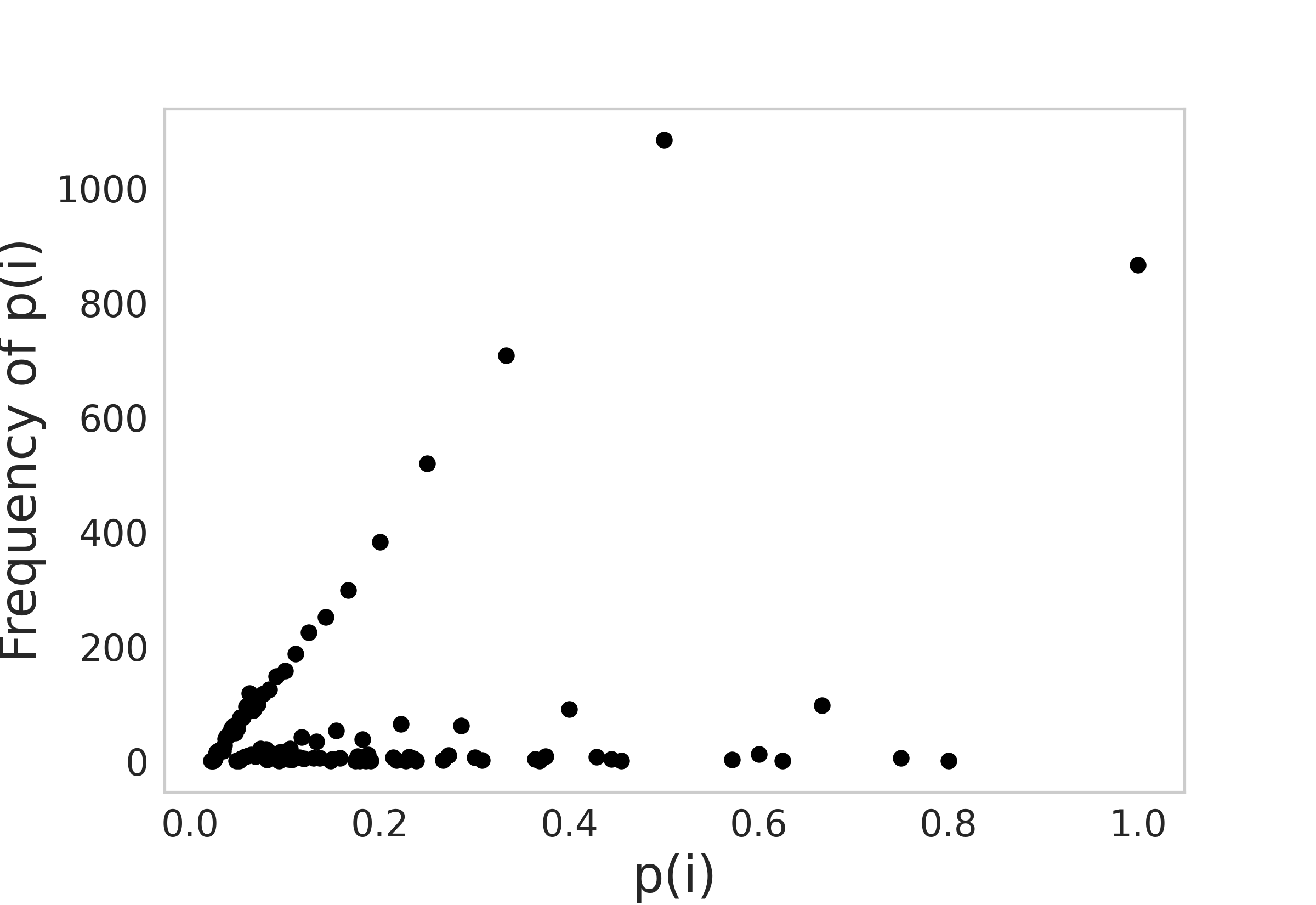}
        \label{fig:MGD}
    }
    \subfloat[in 3-Day Granularity][in 3-Day Granularity]{
        \includegraphics[width=0.5\textwidth]{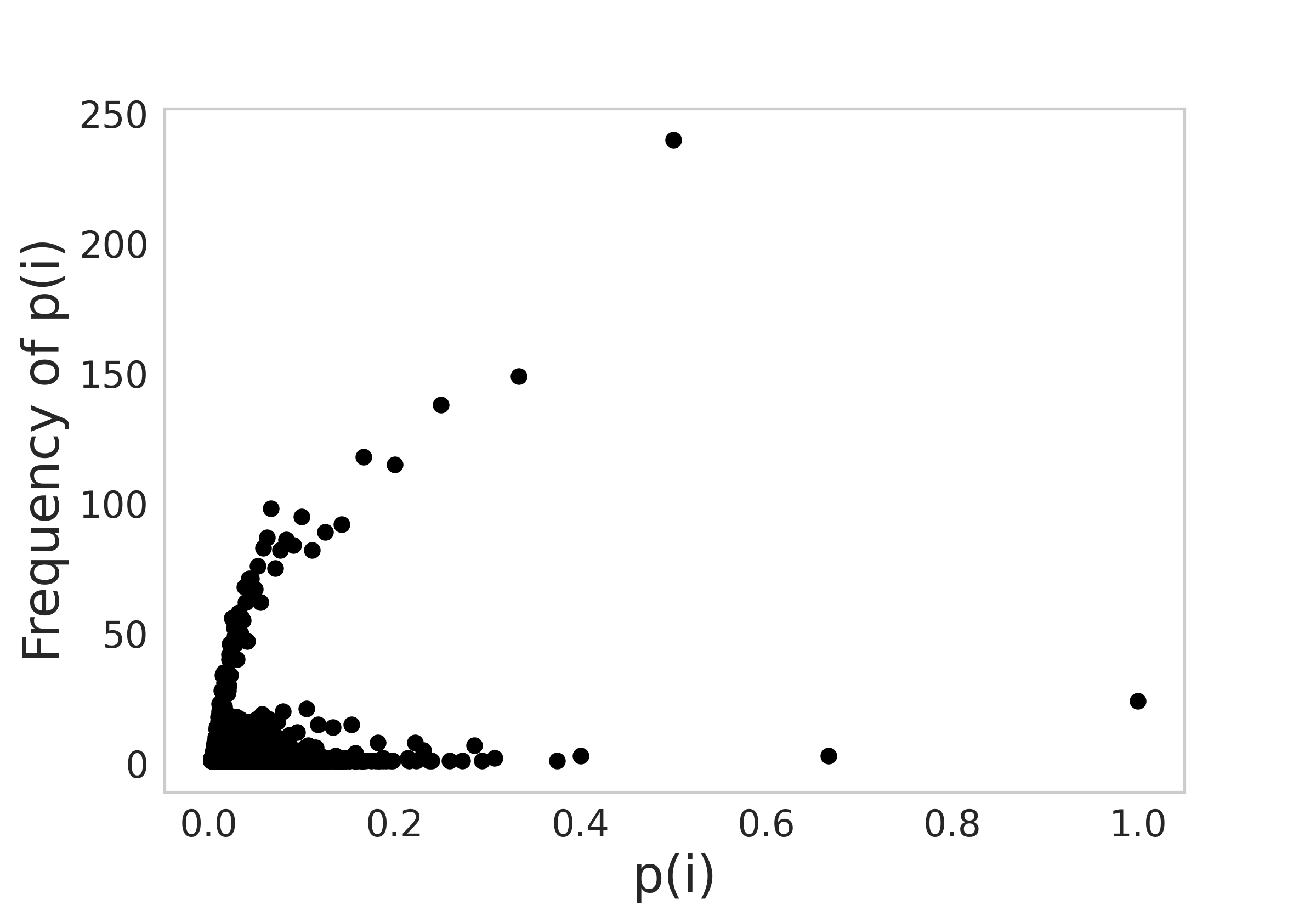}
        \label{fig:TDGD}
    }\\
    \subfloat[in Day Granularity][in Day Granularity]{
        \includegraphics[width=0.5\textwidth]{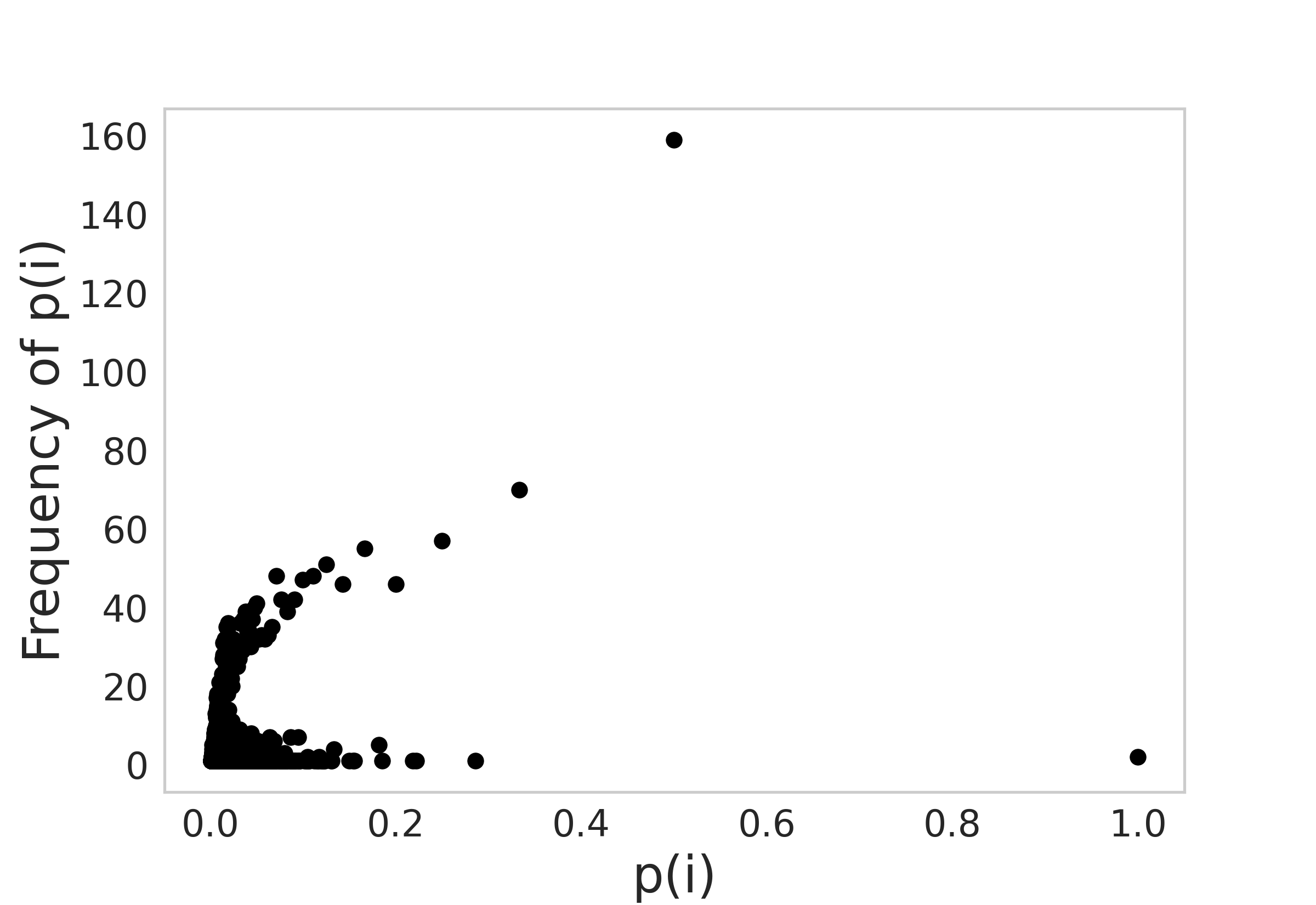}
        \label{fig:DG}
    }
    \caption{Frequency plot of p(i) at different temporal granularities with only transaction features}\label{fig:compositeT}
\end{figure*}

Figures~\ref{fig:compositeTV} and~\ref{fig:compositeT} show the distribution of the frequency of SCs with a particular probability across different temporal granularities with and without considering the severity scores as feature vectors, respectively. Here, we note that the distribution of the frequency of the SCs with a certain probability in 1-Day and 3-Day granularities are similar. This is because the difference between the timeframe represented by these granularities is less. Similarly, the distribution of the frequency of the SCs with a certain probability in the Daywise (1-Day, 3-Day) and 1-Month granularity is different. Again this is because the difference between the timeframe represented by these granularities is high. Also, these suspect SCs (that have a $p=1$) carry out only a few transactions. From these figures, we infer that:

\begin{figure*}
    \centering
    \subfloat[in Month Granularity][in Month Granularity]{
        \includegraphics[width=0.5\textwidth]{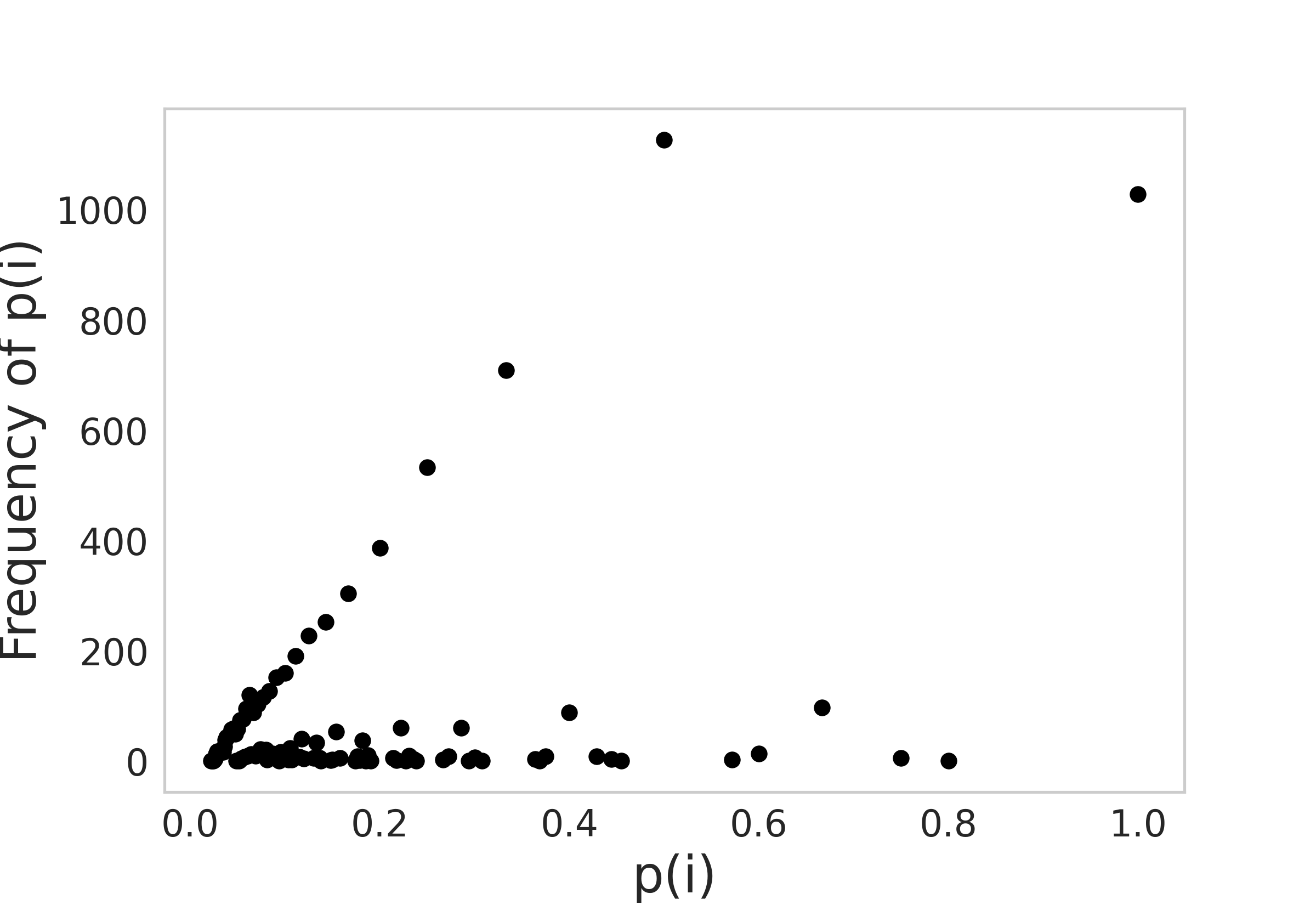}
        \label{fig:MGDS}
    }
    \subfloat[in 3-Day Granularity][in 3-Day Granularity]{
        \includegraphics[width=0.5\textwidth]{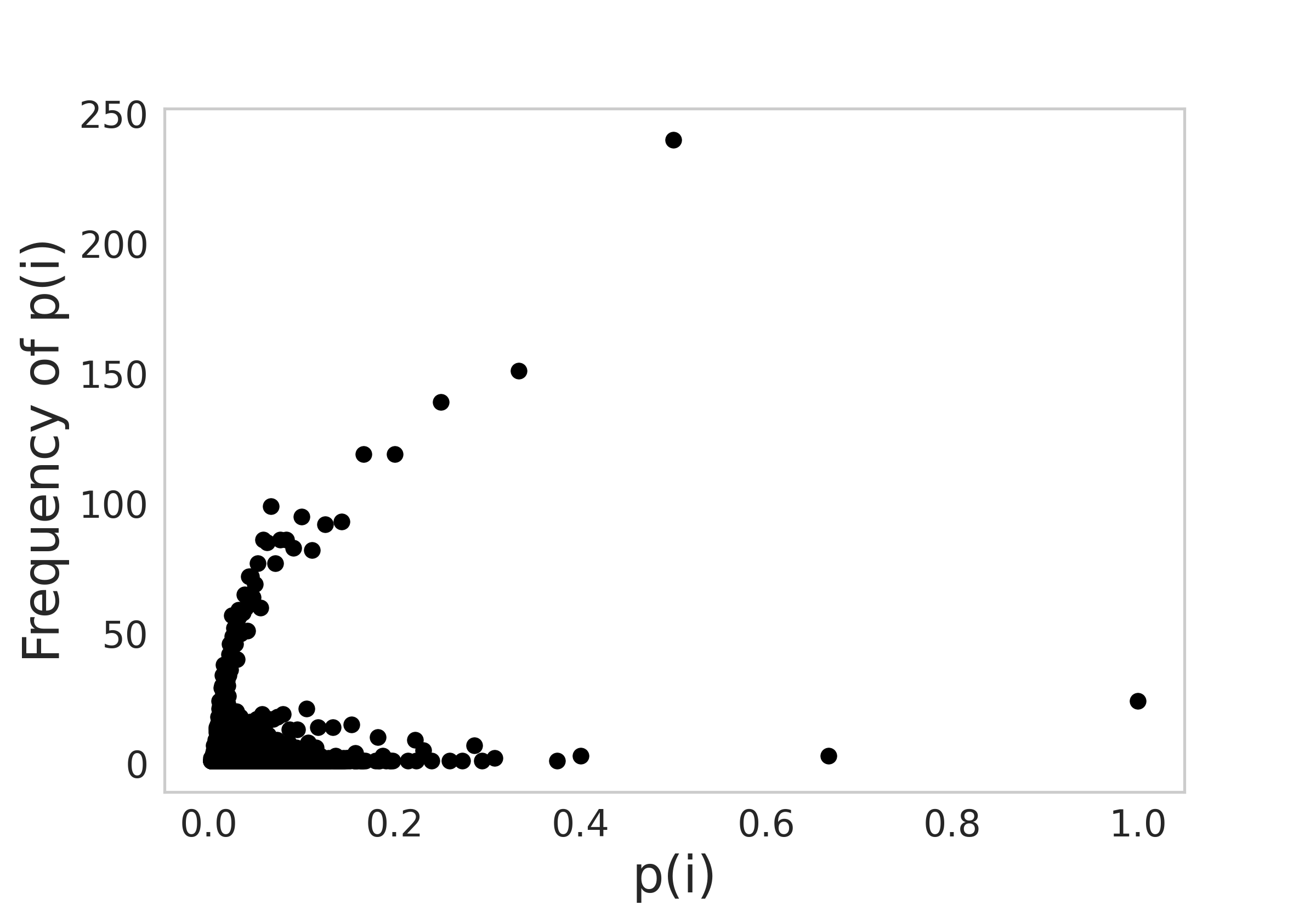}
        \label{fig:TDGDS}
    }\\
    \subfloat[in Day Granularity][in Day Granularity]{
        \includegraphics[width=0.5\textwidth]{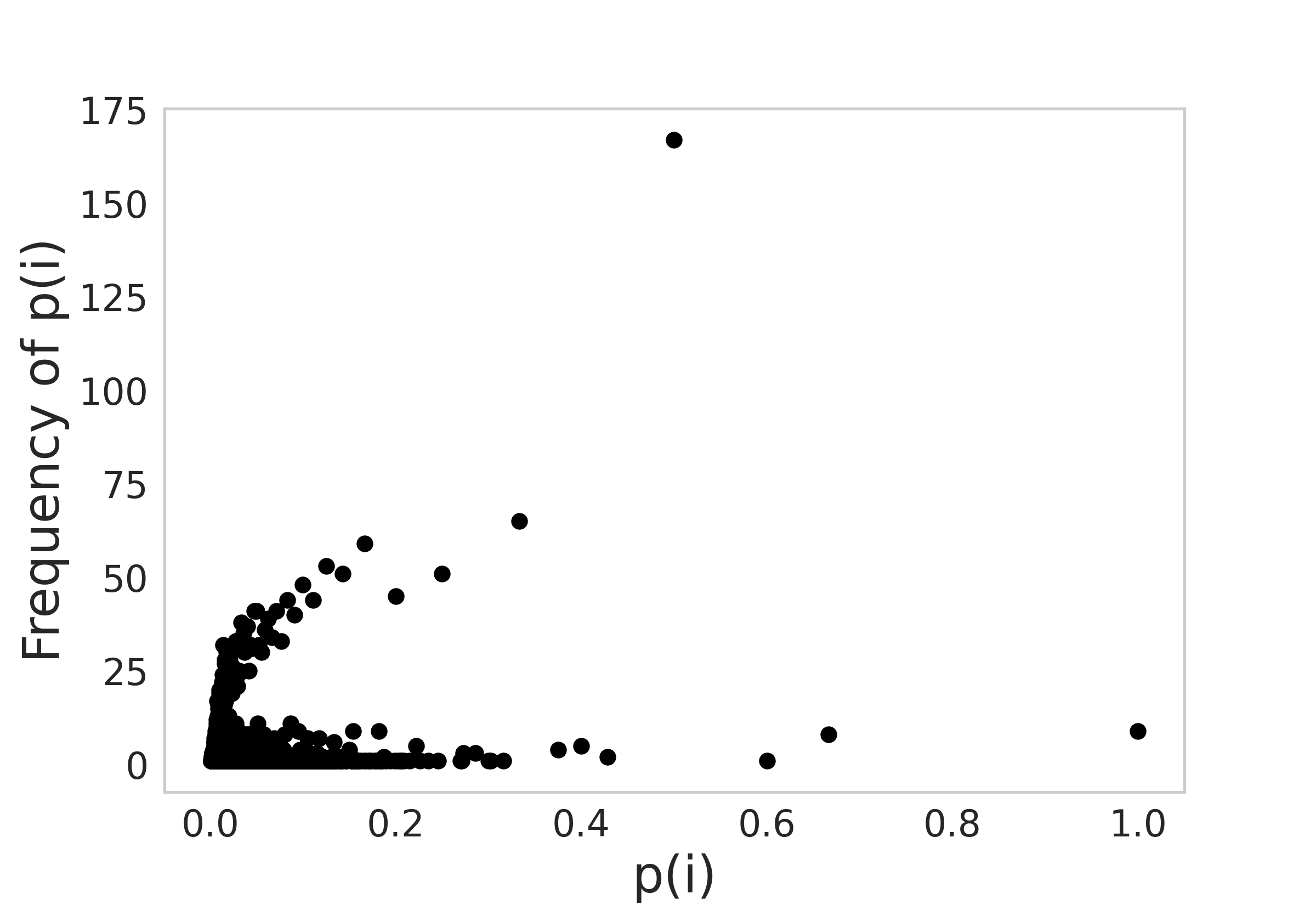}
        \label{fig:DGS}
    }
    \caption{Frequency plot of p(i) at different temporal granularities with both transaction based and severity score as features }\label{fig:compositeTV}
\end{figure*}

\begin{itemize}
\item Most of the SCs have a low probability score. This is represented by the overcrowding of the frequency of the SCs that have low probability scores (cf. Figures~\ref{fig:compositeTV} and~\ref{fig:compositeT}). Note that the low probability score for an SC does not mean that the SC is not malicious. The probability score is less as the number of segments in which the SC was identified as suspects or actually did malicious transaction was less than the number of segments in which the SC carried out the other transactions. 

\item We observe that there are no common suspect SCs between different granularities. Therefore, we infer that the behavior of SCs is changing across different considered temporal granularities.
\end{itemize}

\section{Conclusion and Discussion}\label{sec:conclusion}

The introduction of SCs has opened numerous possibilities for cyber-criminals to steal cryptocurrency and perform illegal activities. Many state-of-the-art approaches leverage ML-based techniques and study transaction behavior to detect accounts held by cyber-criminals. However, these techniques have limitations as they do not distinguish between SCs and other types of accounts (EOAs) in Ethereum. Further, as SCs are programs targeting specific purposes, they have vulnerabilities.   

In this work, we study the correlation between different malicious activities and the vulnerabilities present in SCs. We find that our results are consistent with those of~\cite{Perez2021} as we also do not observe any significant correlation between malicious activities and vulnerabilities. In the process, we also demonstrate the feasibility of using the severity scores of different vulnerabilities as a feature and detect possible suspects amongst the benign SCs. We find that the performance in terms of silhouette score is reduced when we use both the severity and temporal transaction-based features. The severity score feature seems to be a feasible feature for the problem at hand. We also detect different benign suspects across different granularities, such as 1-Day, 3-Day, and 1-Month, using the considered features to understand the temporal behavior changes. Here, we do not get any common suspects across different temporal granularities. This also indicates that the behavior of SCs changes across different considered temporal granularities.  

Note that due to computational restraints, we only considered SCs with unique source codes for our analysis. With more computational resources, it is possible to consider all SCs and their transactions in the study. This could cause changes in the results where some suspicious SCs might occur throughout different temporal granularities. Nonetheless, it may also happen that some SCs have the same vulnerabilities, but their transaction behavior is different. Such aspects would lead to one SC being labeled as malicious while another being benign, which means that the transaction behavior is a more critical factor in identifying malicious SCs than vulnerabilities in the SCs.



\section*{Acknowledgement}\label{sec:ack}
This work is partially funded by the National Blockchain Project (grant number NCSC/CS/2017518) at IIT Kanpur sponsored by the National Cyber Security Coordinator's office of the Government of India and partially by the C3i Center funding from the Science and Engineering Research Board of the Government of India (grant number SERB/CS/2016466). We also thank authors of~\cite{Durieux2020} for providing us their dataset which was partially used in our work. 

\bibliographystyle{IEEEtran}
\bibliography{biblio.bib}

\appendix
\section*{A1. Brief description of DASP top 10 vulnerabilities that are present in our dataset}\label{app:d10}
\begin{itemize}
    \item \textbf{Access Control}: The vulnerabilities which grant an attacker the private variables or functions and logic that are not supposed to be accessible to anyone are grouped under Access Control. One of the many use-cases is when a contract uses the wrong implementation of \textit{tx.origin} for validating SC calls.
    \item \textbf{Arithmetic}: These vulnerabilities are related to the overflows and underflows caused by assigning wrong numeric data types, where the size of the value to be held by the variable is more than the assigned data type's capacity.
    \item \textbf{DoS}: Denial of Service vulnerabilities allows attackers to make an SC lose its functionality. It then cannot provide its `Service' to others. One way of implementing this is to artificially increase the gas necessary to compute a function in an SC.
    \item \textbf{Reentrancy}: according to DASP-10: ``\textit{Reentrancy occurs when external contract calls are allowed to make new calls to the calling contract before the initial execution is complete}". More generally, one interpretation is when a function makes a recursive call. 
    \item \textbf{Unchecked lowlevel call}: Most low-level calls in Solidity do not revert or stop execution if an error is encountered. They return false, and the execution of the function continues. This can lead to unwanted outcomes. For instance, if the sender's SC makes a call to send ether to an SC that doesn't have a payable fallback function and thus doesn't accept them, EVM will replace its return value with false. If the sender does not check this returned Boolean value, then the sender might reduce its own balance by the sent amount, which will not correspond with the actual state of the SC.
    \item \textbf{Transaction Order Dependence (TOD)}: According to DASP-10 ``\textit{Since the Ethereum blockchain is public, everyone can see the contents of others' pending transactions. This means if a given user is revealing the solution to a puzzle or other valuable secret, a malicious user can steal the solution and copy their transaction with higher fees to preempt the original solution.}'' In short, a TOD vulnerability enables an attacker to preempt a transaction by creating another transaction with the solution of the targeted transaction and setting higher gas fees.
    \item \textbf{Timestamp Dependence}: This vulnerability exists when SC uses the block's timestamp to carry out critical operations. Since the miners decide these timestamps, they can manipulate the timestamps to exploit the vulnerability.
    \item \textbf{Bad Randomness}: Some special variables in Ethereum's global namespace have either easy-to-predict values or can be influenced by miners. Suppose an SC uses such variables as a source of randomness. In that case, an attacker(particularly miners) can replicate it and attack any function which uses such variables as a seed in random functions.
    \item \textbf{Short Address Attack}: Such attacks happen when the EVM starts accepting arguments that are not padded correctly. An attacker uses a specially crafted address which causes a client to encode the arguments incorrectly.
    \item \textbf{Unknown Unknowns}: This sub-category comprises vulnerabilities that do not fit in any of the nine sub-categories mentioned above.
    
\end{itemize}

\section*{A2. Feature set}\label{app:fset}
$F=\{$indegreeTimeInv, 
outdegreeTimeInv, 
degreeTimeInv,
numberOfburstTemporalInOut, 
longestBurstTemporalInOut, 
numberOfburstTemporalIn, 
longestBurstTemporalIn, 
numberOfburstTemporalOut, 
longestBurstTemporalOut, 
numberOfburstDegreeInOut, 
longestBurstDegreeInOutAtTime, 
numberOfburstDegreeIn, 
longestBurstDegreeInAtTime, 
numberOfburstDegreeOut, 
longestBurstDegreeOutAtTime, 
zeroTransactions, 
totalBal, 
transactedFirst, 
transactedLast, 
activeDuration, 
averagePerInBal, 
uniqueIn, 
lastActiveSince, 
indegree\_\_index\_mass\_quantile\_\_q\_0.1,
indegree\_\_energy\_ratio\_by\_chunks\_\_num\_segments\_10\_\_segment\_focus\_0, 
indegree\_\_linear\_trend\_\_attr\_``pvalue", 
ittime\_\_quantile\_\_q\_0.7, 
ittime\_\_fft\_coefficient\_\_coeff\_0\_\_attr\_``real", 
ittime \_\_median, 
outdegree\_\_energy\_ratio\_by\_ chunks\_\_num\_segments\_10\_\_segment\_focus\_0, 
outdegree\_\_enegy\_ratio\_by\_chunks\_\_num\_segments\_10\_\_segment\_ focus\_1, 
outdegree\_\_fft\_coefficient\_\_coeff\_0\_\_attr\_``real", 
gasPrice\_\_quantile \_\_q\_0.2, 
gasPrice\_\_quantile\_\_q\_0.1, 
gas-Price\_\_cwt\_coefficients\_\_widths\_(2, 5, 10, 20)\_\_coeff\_0\_\_w\_20, 
attractiveness\_\_median, 
attractiveness\_\_quantile\_\_q\_0\_ 0.4, 
attractiveness\_\_mean, 
balanceOut\_\_quantile\_\_q\_0.1, 
balanceOut\_\_quantile\_\_q\_0.3, 
balanceOut\_\_cwt\_coefficients\_\_ widths\_(2, 5, 10, 20)\_\_coeff\_0\_\_w\_2, 
balanceIn\_\_quantile \_\_q\_0.4, 
balanceIn\_\_cwt\_coefficients\_\_widths\_(2, 5, 10, 20)\_\_coeff\_0\_\_w\_20, 
balanceIn\_\_quantile\_\_q\_0.3, 
maxInPayment\_\_quantile\_\_q\_0.3, 
maxInPayment\_\_quantile\_\_q\_0.2, 
maxInPayment\_\_cwt\_coefficients\_\_widths\_(2, 5, 10, 20)\_\_coeff\_0\_\_w\_5, 
maxOutPayment\_\_quantile\_\_q\_0.6, 
maxOutPayment\_\_quantile\_\_q\_0.1, 
maxOutPayment \_\_cwt\_coefficients\_\_widths\_(2, 5, 10, 20)\_\_coeff\_0\_\_w\_2, 
clusteringCoeff, 
burstCount\_gasPrice, 
burstCount\_balanceIn, 
burstCount\_balanceOut, 
burstInstance\_indegree, 
burstInstance\_outdegree, 
burstInstance\_outdegree, 
burstInstance\_maxInPayment, 
burstInstance\_maxOutPayment, 
burstInstance\_gasPrice$\}$

Note that the features  identified using time series analysis are succeeded by `\_\_' which is further succeeded by the parameter name and their values. 

\end{document}